\documentclass[11pt,letter,notitlepage]{article}
\usepackage[left=1 in, top=1 in, right=1 in, bottom=1 in]{geometry}

\usepackage{graphicx}
\usepackage{amsmath}
\usepackage{amsfonts}
\usepackage{amssymb}
\usepackage{multicol}
\usepackage{fancyhdr}
\usepackage{cases}
\usepackage{sectsty} 
\usepackage{tocloft} 
\usepackage{hyperref}
\usepackage{url}
\usepackage{authblk}
\usepackage[left]{lineno}
\linenumbers

\usepackage{setspace}

\usepackage{enumitem}
\setlist[itemize]{noitemsep,topsep=0pt}

\usepackage[utf8]{inputenc} 
\usepackage[backend=bibtex,style=nature]{biblatex}
\parskip = \baselineskip

\subsectionfont{\normalfont\itshape} 

\newcommand{\w}{_\mathrm{\textsc{w}}}
\newcommand{\m}{_\mathrm{\textsc{m}}}

\newcommand{\beginsupplement}{
        \setcounter{table}{0}
        \renewcommand{\thetable}{S\arabic{table}}
        \setcounter{figure}{0}
        \renewcommand{\thefigure}{S\arabic{figure}}
        \setcounter{equation}{0}
        \renewcommand{\theequation}{S\arabic{equation}}
} 
\vspace{-3in}

\title{\Large \textbf{Slow expanders invade by forming dented fronts in microbial colonies}}

\author[1]{Hyunseok Lee}
\author[1]{Jeff Gore}
\author[2]{Kirill S. Korolev\thanks{\nolinkurl{korolev@bu.edu}}}
\affil[1]{Physics of Living Systems Group, Department of Physics, Massachusetts Institute of Technology, Cambridge, MA 02139}
\affil[2]{Department of Physics, Graduate Program in Bioinformatics, Biological Design Center, Boston University, Boston, MA 02215}

\date{}

\begin{document}

\maketitle

\begin{abstract}
\noindent\textbf{Abstract: }\\
Most organisms grow in space, whether they are viruses spreading within a host tissue or invasive species colonizing a new continent. Evolution typically selects for higher expansion rates during spatial growth, but it has been suggested that slower expanders can take over under certain conditions. Here, we report an experimental observation of such population dynamics. We demonstrate that the slower mutants win not only when the two types are intermixed at the front but also when they are spatially segregated into sectors. The latter was thought to be impossible because previous studies focused exclusively on the global competitions mediated by expansion velocities but overlooked the local competitions at sector boundaries. We developed a theory of sector geometry that accounts for both local and global competitions and describes all possible sector shapes. In particular, the theory predicted that a slower, but more competitive, mutant forms a dented V-shaped sector as it takes over the expansion front. Such sectors were indeed observed experimentally and their shapes matched up quantitatively with the theory. In simulations, we further explored several mechanism that could provide slow expanders with a local competitive advantage and showed that they are all well-described by our theory. Taken together, our results shed light on previously unexplored outcomes of spatial competition and establish a universal framework to understand evolutionary and ecological dynamics in expanding populations.
\end{abstract}

\section*{Significance}
Living organisms never cease to evolve, so there is a significant interest in predicting and controlling evolution in all branches of life sciences from medicine to agriculture. The most basic question is whether a trait should increase or decrease in a given environment. The answer seems to be trivial for traits such as the growth rate in a bioreactor or the expansion rate of a tumor. Yet, it has been suggested that such traits can decrease rather than increase during evolution. Here, we report a mutant that outcompeted the ancestor despite having a slower expansion velocity. To explain this observation, we developed and validated a theory that describes spatial competition between organisms with different expansion rates and arbitrary competitive interactions.

\section*{Introduction}
Population dynamics always unfold in a physical space. At small scales, microbes form tight associations with each other, substrates, or host cells~\cite{grossart2003bacterial, datta2016microbial}. At large scales, phyto- and zooplanktons form complex patterns influenced by ecological interactions~\cite{durrett1998spatial, ben2000cooperative, lima2015determinants} and hydrodynamics~\cite{abraham1998generation, pigolotti2012population}. Between these two extremes, populations constantly shrink and expand in response to changing conditions, and there is still a great deal to be learned about how spatial structure affects ecology and evolution~\cite{skellam1951random, levin1974dispersion,  hanski1998metapopulation, lin2015demographic, nadell2016spatial}. Better understanding of these eco-evolutionary dynamics is essential for management of invasive species~\cite{mooney2001evolutionary, sakai2001population}, controlling the growth of cancer~\cite{korolev2014turning}, and preserving biodiversity~\cite{hastings2005spatial, jeschke2018invasion}.

It is particularly important to understand how natural selection operates at the edge of expanding populations. These expansion frontiers are hot spots of evolution because mutations that arise at the edge can rapidly establish over large areas via allele surfing or sectoring~\cite{barton1984genetic, klopfstein2006fate, hallatschek2008gene, hallatschek2014acceleration}.  Furthermore, numerous studies argue that selection at the expansion front favors faster expanders and therefore makes population control more difficult~\cite{kot1996dispersal, thomas2001ecological, benichou2012front, shine2011evolutionary, van2013convergent, korolev2013fate, yi2016phenotypic, fraebel2017environment, ni2017evolutionary, shih2018biophysical, deforet2019evolution}. Indeed, organisms that expand faster have a head start on growing into a new territory and may face weaker competition or better access to nutrients. A well-known example is the evolution of cane toads which increased the expansion speed by 5 fold over 50 years~\cite{phillips2006invasion}. Yet, despite substantial empirical evidence across many systems~\cite{thomas2001ecological,shine2011evolutionary,benichou2012front,van2013convergent,yi2016phenotypic,fraebel2017environment,ni2017evolutionary,shih2018biophysical,deforet2019evolution}, it has been suggested that the simple intuition of ``faster runner wins the race'' does not always hold.

Two theoretical studies have found that slower dispersal could evolve in populations with a strong Allee effect, i.e a negative growth rate at low population densities~\cite{travis2002dispersal, taylor2005allee, korolev2015evolution}. Slow mutants nevertheless can take over the populations because they are less likely to disperse ahead of the front into regions with low densities and negative growth rates. In a different context, both theory and experiments have shown that slow cheaters could invade the growth front of fast cooperators~\cite{korolev2013fate, datta2013range}. In this system, the production of public goods allowed cooperators to expand faster, but made them vulnerable to the invasion by cheaters. 

The examples above show that slower expanders succeed in the presence of a tradeoff between local and global fitness. The global fitness is simply the expansion rate of a given species in isolation, which determines how quickly it can colonize an empty territory. When two species are well-separated in space, their competition is determined solely by the global fitness. In contrast, when the two species are present at the same location, their competition could involve differences in growth rates, production of public goods~\cite{allen2013spatial, bauer2018multiple}, or secretion of toxins~\cite{stempler2017interspecies}. We refer to such local competitive abilities as local fitness. It is natural to assume that slow expanders can win only if they are superior local competitors, but it is not clear a priori if this is actually feasible or how to integrate local and global fitness under various scenarios of spatial competition.

Our interest in the interplay between local and global competition was sparked by an unusual spatial pattern in colonies of \textit{Raoultella planticola} grown on agar plates. These colonies repeatedly developed depressions or dents along the edge. We found that dents were produced by a spontaneous mutant that expanded slower than the wildtype. Thus, we discovered a convenient platform to explore the fate of slower expanders in spatial competition and to elucidate the tension between local and global fitness.

In our experiment, the slower expander took over the colony either by increasing in frequency homogeneously along the front or by forming pure, mutant-only, sectors.  When mutant sectors formed, they had an unusual ``dented'' or ``V'' shape. To explain this spatial pattern, we developed a theory that describes all possible sector geometries. Our theory unifies local and global competitions without assuming any particular mechanism for growth and dispersal. Although mechanism-free, the theory makes quantitative predictions, which we confirmed experimentally. We also simulated multiple mechanistic models to demonstrate that the takeover by slower expanders is generic and could occur due to multiple ecological mechanisms. These simulations further confirmed that sector shape prediction from geometric theory is universal. Taken together, our results establish a new framework to understand evolutionary and ecological dynamics in expanding populations with arbitrary frequency- and density-dependent selection.

\section*{Results}
\subsection*{Experimental observation of slow mutants taking over the front}
The strains used in our experiment were derived from a soil isolate of \textit{Raoultella planticola}, a Gram-negative, facultatively anaerobic, non-motile bacterium that is found in soil and water and can occasionally lead to infections~\cite{drancourt2001phylogenetic, ershadi2014emerging}. We grew \textit{R. planticola} on a hard LB agar plate ($1.5\%$ agar) and noticed the formation of V-shaped dents along the front. Such dents were reproducibly observed in biological replicates~(Fig.~\ref{fig:si1}). Suspecting that dents were caused by a mutation, we isolated cells from the smooth parts of the colony edge~(wildtype) and from the dents~(mutant)~(Fig.~\ref{fig:1}A).  
 
\begin{figure}
\centering
\includegraphics[width=\textwidth]{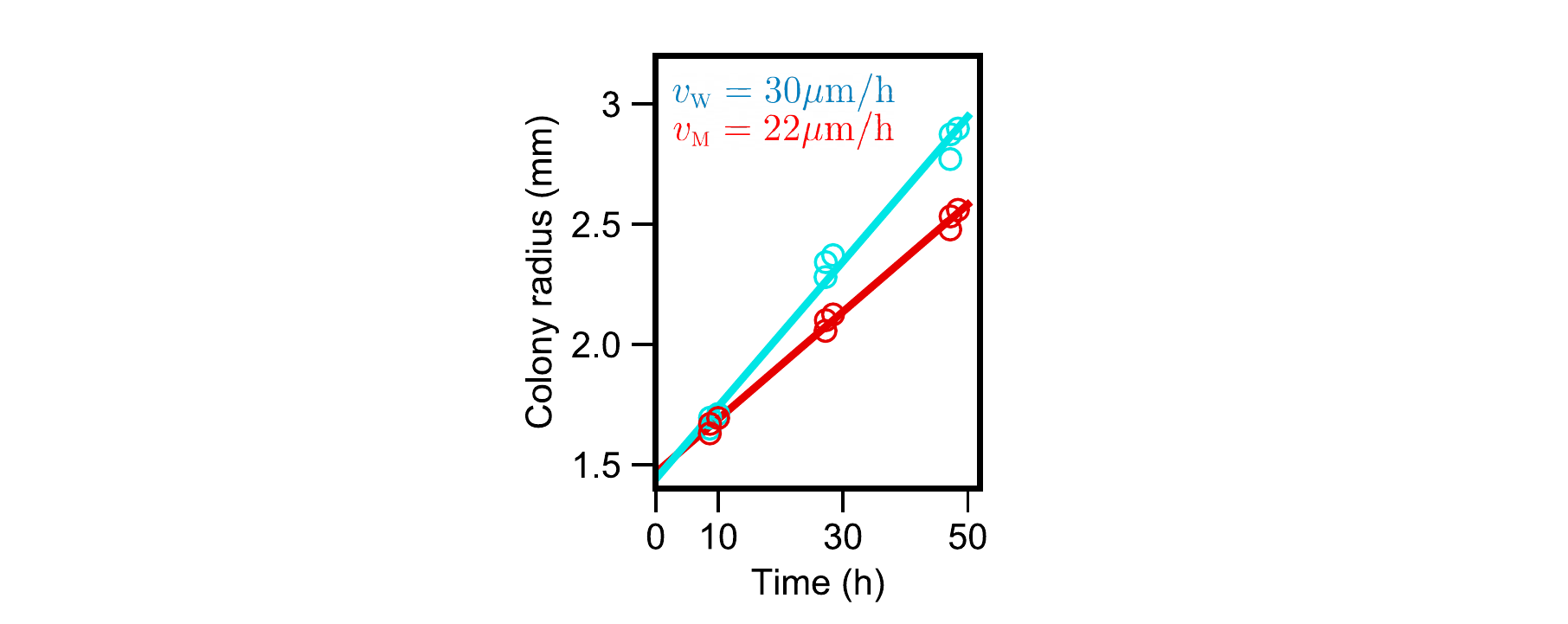}
\caption{Mutant expands more slowly regardless of the choice of fluorescent labels. Wildtype with mTurquoise-2 fluorescence protein expanded with $v\w=30\;\mu m/h$ while mutant with mScarlet-I fluorescence protein expanded with $v\m = 22\; \mu m/h$.  }\label{fig:si2}
\end{figure}

We first characterized the expansion dynamics of the two strains in isolation by inoculating each culture at the center of a hard agar plate. Both strains formed smooth, round colonies, which expanded at a constant velocity~(Fig.~\ref{fig:1}B, Fig.~\ref{fig:si2}). The wildtype had about~50\% larger expansion velocity compared to the mutant. Thus, the evolved strain was a slower expander. 

Our observations seemed paradoxical given numerous observations of invasion acceleration due to genetic changes that increase expansion velocities~\cite{simmons2004changes, phillips2006invasion}. However, range expansions are known to produce high genetic drift~\cite{waters2013founder, birzu2019genetic} and, therefore, allow for the fixation of deleterious mutations~\cite{hallatschek2008gene, excoffier2009genetic, roques2012allee, slatkin2012serial, bosshard2017accumulation}. So, we next investigated whether the mutant has a selective advantage in competition with the wildtype within the same colony.

\begin{figure}
\centering
\includegraphics[width=\textwidth]{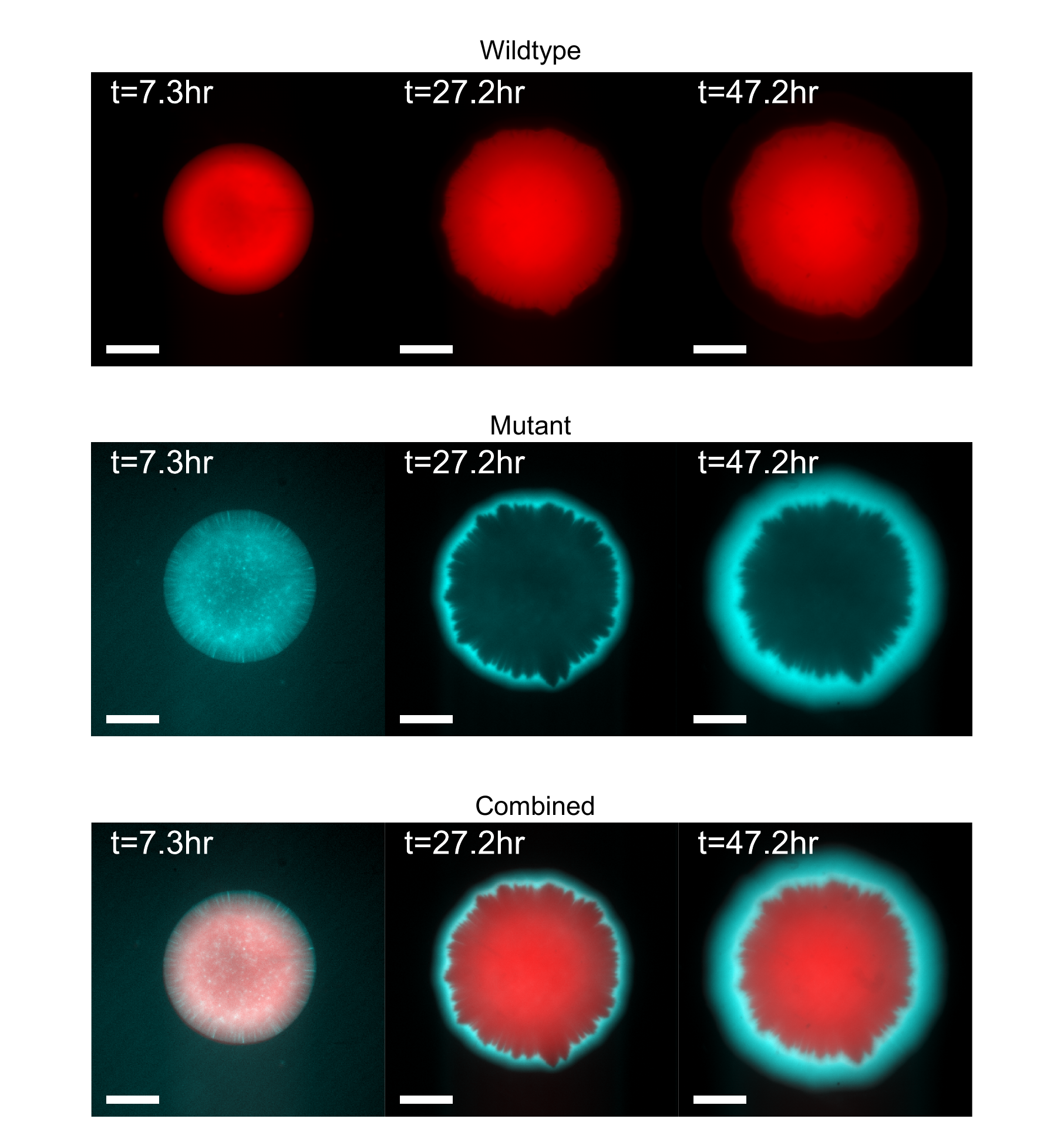}
\caption{In co-culture experiment, wildtype did not expand after a day while mutant kept expanding. Top: Fluorescence images of wildtype cells during expansion. Middle: Fluorescence images of mutant cells during expansion. Bottom: combined. }\label{fig:si3}
\end{figure}

We competed the two strains by inoculating an agar plate with a drop containing a 99:1 mixture of the wildtype and the mutant. We used two wildtype strains (and their respective mutants) with different fluorescent labels and the spatial patterns were analyzed with fluorescence microscopy (see Methods). After about 48 hours of growth, a ring of mutant completely encircled the wildtype~(Fig.~\ref{fig:1}C). Only the mutant ring continued to expand, while the expansion of the wildtype ceased~(Fig.~\ref{fig:si3}). Thus the mutant not only localized to the front but also achieved a greater population size. This is quite different from other microbial systems where a strain with poor motility localized to the front without suppressing the growth of faster strain and without producing a larger biomass~\cite{yan2017extracellular, xiong2020flower}. Thus, our experiments strongly suggest that the mutant has a competitive advantage despite its lower expansion velocity.

\subsection*{Experimental observation of slow mutants invading by forming dented fronts}

\begin{figure}[htp]
    \centering
    \includegraphics[width=.8\linewidth]{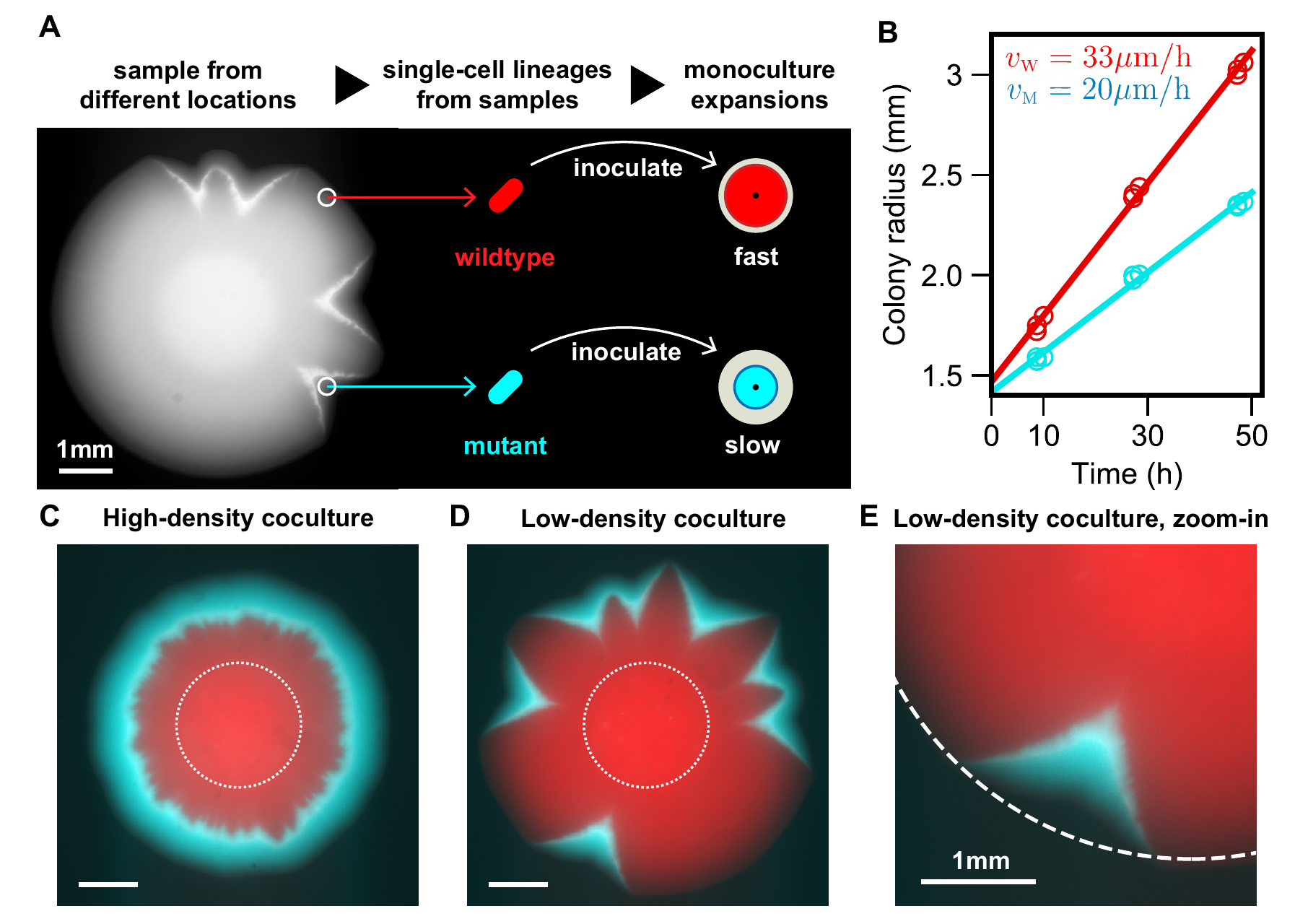}
    \caption{Slow mutant takes over the front with and without sector formation. (A) We found that wildtype \textit{R. planticola} colonies develop V-shaped indentations; a bright-field image is shown. We sampled cells from the dents and non-dented regions and then developed strains descending from a single cell~(see Methods). (B)~The mutant expanded more slowly than the wildtype. The data points come from two technical replicates, and the line is a fit. (C, D)~Despite its slower expansion, the mutant wins in coculture. Fluorescence images show the spatial patterns 48~hours after inoculation with a 99:1 mixture of the wildtype and mutant. A ring of mutant (cyan) outrun and encircled wildtype (red) when the mixed inoculant had a high density ($\mathrm{OD_{600}}$ of~$10^{-1}$). Mutant sectors emerged and widened over the front when the mixed inoculant had a low density ($\mathrm{OD_{600}}$ of~$10^{-3}$). Images are taken 48~hours after inoculation, and dotted lines represent initial inoculant droplets. (E) A zoomed image of a V-shaped sector~(from the bottom of D). Dotted circle is a fit from wildtype expansion. The advantage of the mutant and its slower expansion is evident from the lateral expansion of the cyan sector.}
    \label{fig:1}
\end{figure}

Our initial competition experiments did not exhibit the dents that sparked our initial interest in the strains. The mutant took over uniformly across the expansion front, producing a rotationally invariant spatial pattern~(Fig.~\ref{fig:1}C). In fact, one might even argue that the success of the mutant could have been entirely due to the transient growth dynamics, and the wildtype would prevail if allowed to somehow spatially segregate from the mutant. To address both of these concerns, we sought to alter the experiments so that the mutant and the wildtype grow as distinct sectors within the same colony.

In microbial colonies, sectors emerge due to genetic drift at the growing edge. The magnitude of demographic fluctuations varies widely in different systems, depending on the organism, the growth conditions, and the duration of the experiment~\cite{hallatschek2007genetic, korolevRMP2010}. To test for the effects of sectoring, we needed to increase stochasticity without altering other aspects of the competition. Reducing the cell density of the initial inoculant accomplished this goal. By lowering the inoculant density (from~$10^{-1}$ $\mathrm{OD_{600}}$ to~$10^{-3}$ $\mathrm{OD_{600}}$), we increased the separation between cells that localized to the colony edge following the drying of the inoculation drop. This in turn dramatically increased the formation of monoclonal sectors~(Fig.~\ref{fig:1}D).  

\begin{figure}
\centering
\includegraphics[width=\textwidth]{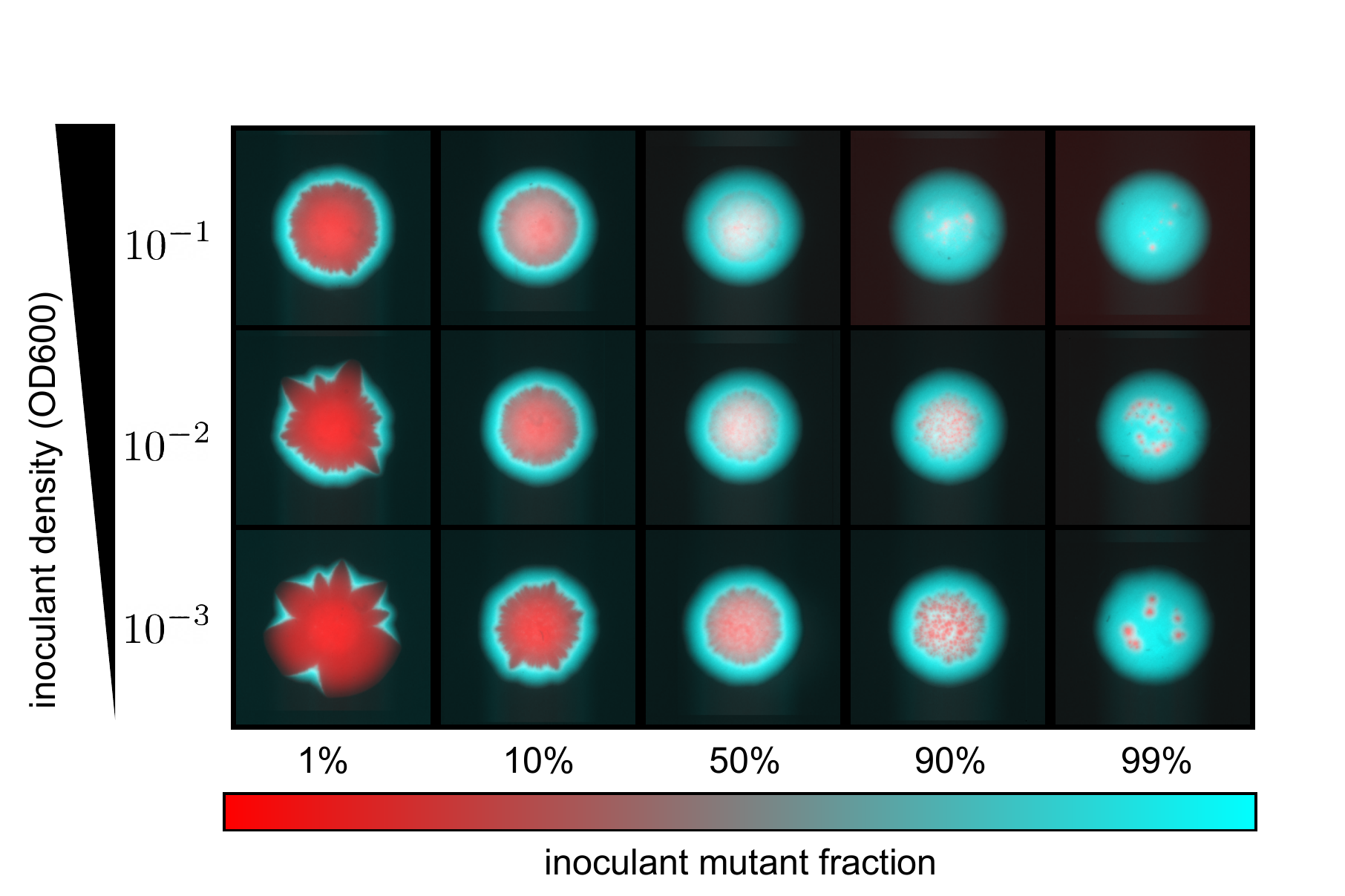}
\caption{Mutant outcompetes wildtype under a wide range of inoculant densities and initial mutant fractions. }\label{fig:si4}
\end{figure}

\begin{figure}
\centering
\includegraphics[width=\textwidth]{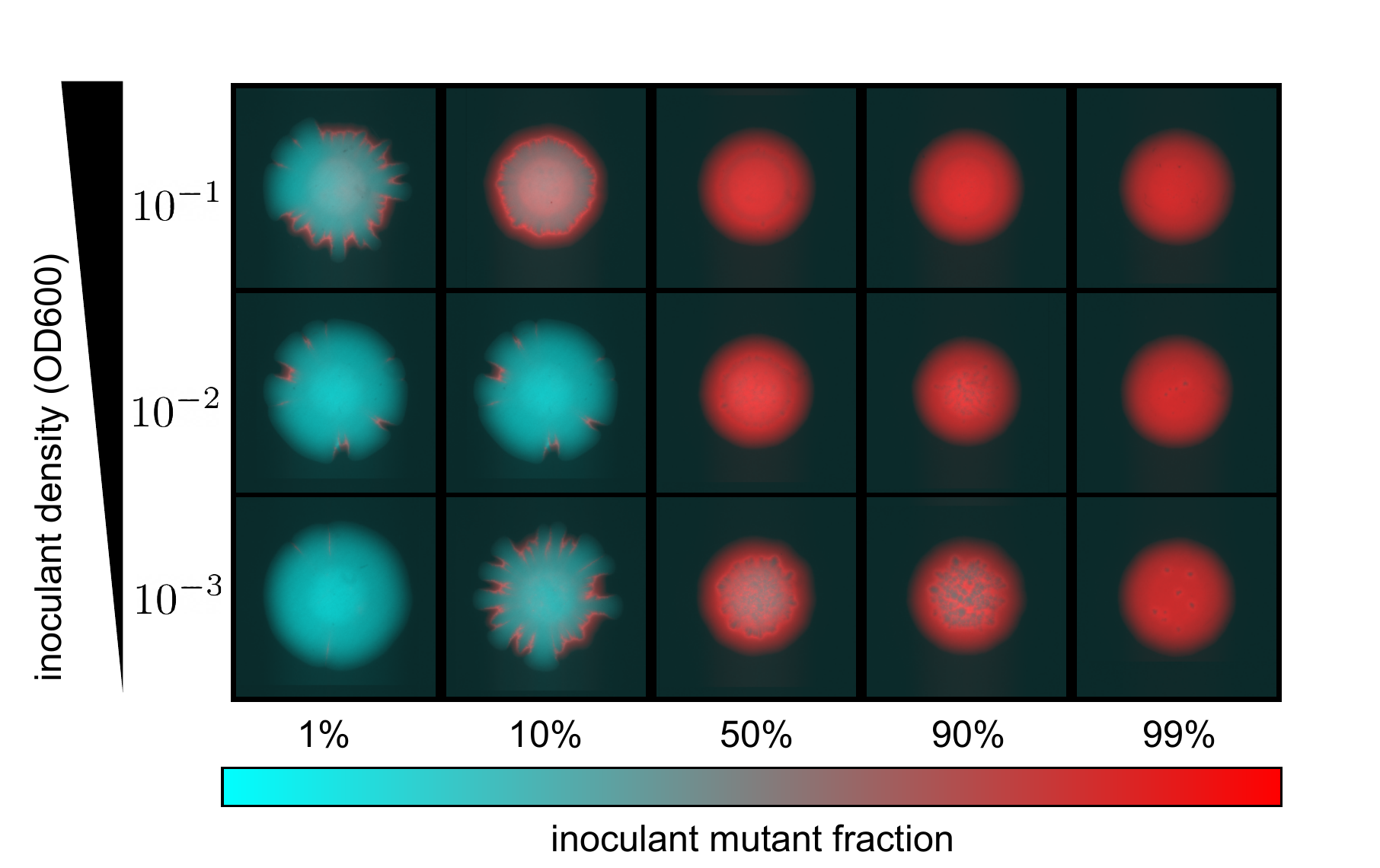}
\caption{Mutant outcompetes wildtype under a different choice of fluorescent labels of wildtype and mutant. }\label{fig:si5}
\end{figure}

Although sectoring spatially segregated the two strains and, thus, allowed the wildtype to expand with a higher velocity, the slower mutant still outcompeted the wildtype~(Fig.~\ref{fig:1}D, Fig.~\ref{fig:1}E). The takeover of mutant was robust under different choices of initial density, initial mutant fraction, and fluorescent label~(Fig.~\ref{fig:si4}, Fig.~\ref{fig:si5}). The takeover by the mutant also produced the characteristic V-shaped dents at the colony edge. These dents are the exact opposite of the bulges or protrusions that one usually observes for beneficial mutations~\cite{korolev:sectors}. Typically,  the advantageous mutants have a greater expansion velocity and, therefore, outgrow the ancestors at the front. For our strains, however, the winning mutant had a lower expansion velocity, and this lower expansion velocity produced the opposite of the bulge---the dent.

\subsection*{Mechanism-free theory of sector geometry}

Our experiments unambiguously demonstrated that a slower expander can indeed outcompete a faster expander with and without sectoring. Still, we need a careful theoretical description of the spatial dynamics to reconcile the apparent contradiction between the slow global expansion of the mutant and its superior performance in local competition. We could approach this question by simulating a specific ecological mechanism that could be responsible for the tradeoff between local and global fitness. However, it is much more useful to first ask what can be said about spatial competition generically and determine the range of possible sector shapes without relying on any specific mechanism.

Our analysis follows the approach similar to geometric optics in physics~\cite{bressan2007differential, horowitz2019bacterial, lipson2010optical} and it relies on a few standard assumptions. The expansion velocities of the two strains~($v_{\mathrm{\textsc{w}}}$ and~$v_{\mathrm{\textsc{m}}}$) are assumed to be time-independent both to simplify the calculations and to reflect experimental observation~(Fig.~\ref{fig:1}B). We also assume, consistent with past studies~\cite{korolevAmericanNaturalist, muller2014genetic, momeni2013strong}, that there is little growth behind the front so that the spatial pattern remains once established as in our experiments. Finally, we neglect long-range interactions due to the diffusion of nutrients, toxins, or signaling molecules\footnote{The addition of long-range interactions would provide greater modelling flexibility and therefore make it easier to observe novel spatial patterns such as a V-shaped sector. Our works shows that this extra flexibility is unnecessary and dented fronts can appear in purely local models.}~\cite{prindle2015ion, mitri2016resource, cremer2019chemotaxis}. 

\begin{figure}[htp]
    \centering
    \includegraphics[width=.8\linewidth]{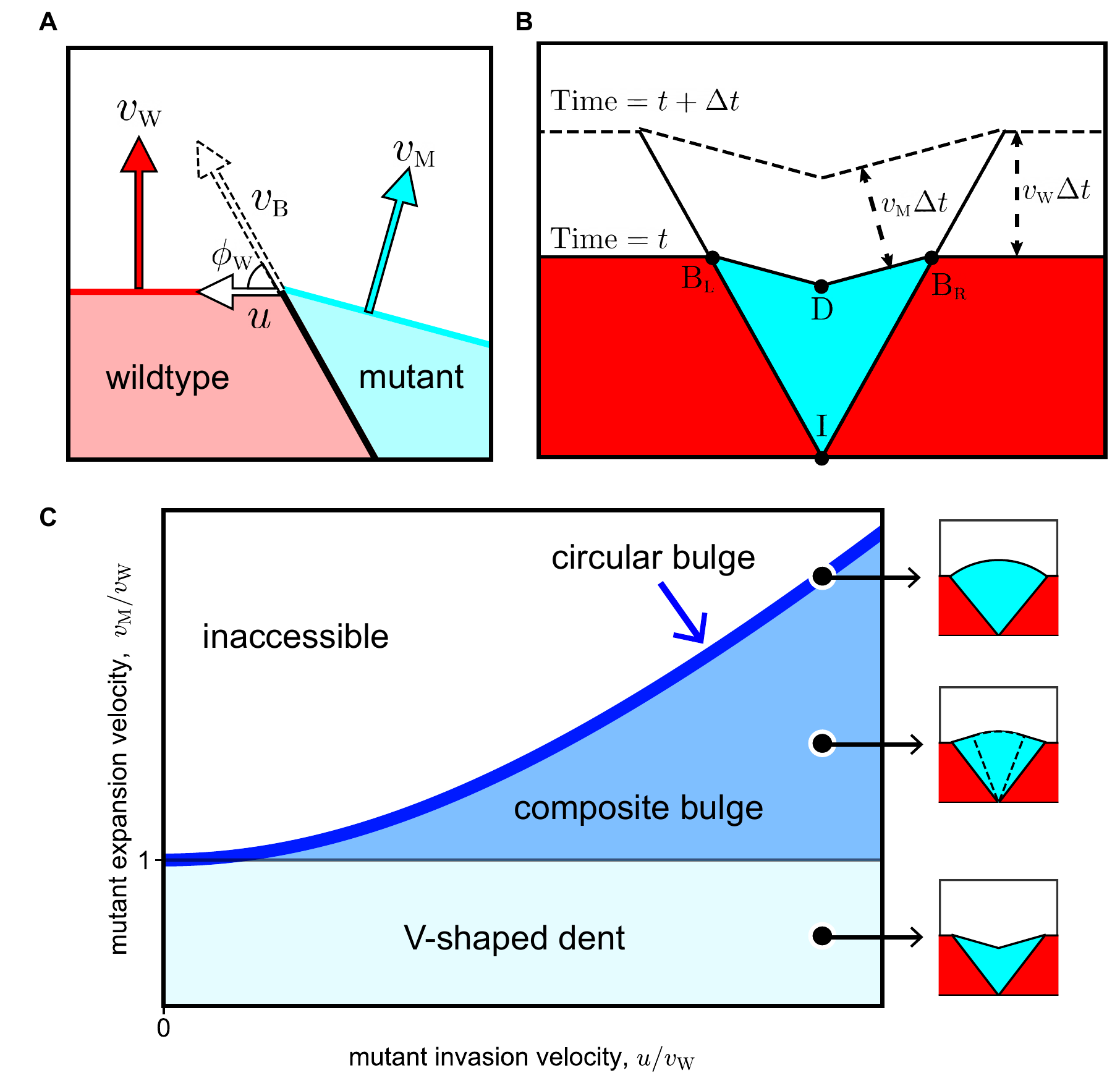}
    \caption{Geometric theory predicts sector shapes as a function of local and global fitness. Flat-front initial conditions are illustrated here, and the corresponding results for circular fronts are shown in the SI. (A)~Global fitness of mutant and wildtype are defined with speeds $v\w$ and $v\m$ with which their fronts advance. Local fitness is defined by movement of sector boundary which advances with speed $v_{\mathrm{\textsc{b}}}$ at angle $\phi\w$ with the wildtype front. We use lateral invasion speed $u$ to describe the local fitness, and the equivalence among $v_{\mathrm{\textsc{b}}}$, $\phi\w$, and $u$ is explained in SI~\ref{sec:geometric}. (B)~The shape of the mutant sector can be derived from geometric considerations. During a time interval~$\Delta t$, the boundary points~$\mathrm{B}_\mathrm{\textsc{l}}$ and~$\mathrm{B}_\mathrm{\textsc{r}}$ move upward by~$v\w \Delta t$ and laterally outward by~$u \Delta t$. The position of the dent~$\mathrm{D}$ is obtained from the requirement that both~$\overline{\mathrm{DB}_\mathrm{\textsc{l}}}$ and~$\overline{\mathrm{DB}_\mathrm{\textsc{r}}}$ shift by~$v\m \Delta t$; the directions of the shifts are perpendicular to~$\overline{\mathrm{DB}_\mathrm{\textsc{l}}}$ and~$\overline{\mathrm{DB}_\mathrm{\textsc{r}}}$ respectively. Point~$\mathrm{I}$ labels the origin of the sector. 
     (C)~The geometric theory predicts sector shapes as a function of~$u/v\w$ and~$v\m/v\w$.  When~$v\m < v\w$ and~$u>0$, the mutant forms a V-shaped dented front; note that all boundaries are straight lines. When~$v\m > v\w$ and~$u > \sqrt{v^2\m - v^2\w}$, the mutant forms a bulged front. The shape of the bulge consists of two regions. It is an arc of a circle near the middle and two straight lines near the two boundaries between the mutant and the wildtype. The circular region grows and the linear region shrinks as $v{\m}/v{\w}$ increases at constant~$u/v{\w}$. The bulge becomes completely circular when~$v\m/v{\w}$ reaches its maximal value of~$\sqrt{1+u^2/v_{\w}^2}$ on the boundary of the accessible region. See SI for derivation and exact mathematical expressions of all sector shapes. }
    \label{fig:2}
\end{figure}

The nontrivial aspect of our work is how we capture the effect of local competition between strains. This can be done in a number of equivalent ways. The most intuitive one is to define a velocity $u$ with which the mutant invades laterally into the population of the wildtype. Alternatively, one can consider the velocity of the boundary between the strains: $v_{\mathrm{\textsc{b}}}$, which cannot be inferred solely from $v\w$ and $v\m$ and thus contains information about local fitness. and the angles between the boundary and expansion fronts. The connection between these approaches is illustrated in Fig.~\ref{fig:2}A. 

The knowledge of the three velocities~($v_{\mathrm{\textsc{w}}}$,~$v_{\mathrm{\textsc{m}}}$, and~$u$) is sufficient to simulate how the shape of the colony changes with time. In some situations, colony shapes can also be obtained analytically by comparing the position of the front at two times~$t$ and~$t+\Delta t$. We derive the equations for sector shapes by requiring that all distances between the corresponding points of the two fronts are given by~$\Delta t$ times the appropriate velocity~(Fig.~\ref{fig:2}B). The details of these calculations are provided in the SI~(Fig.~\ref{fig:g2}).

\begin{figure}[ht!]
	\includegraphics[width=0.8\linewidth]{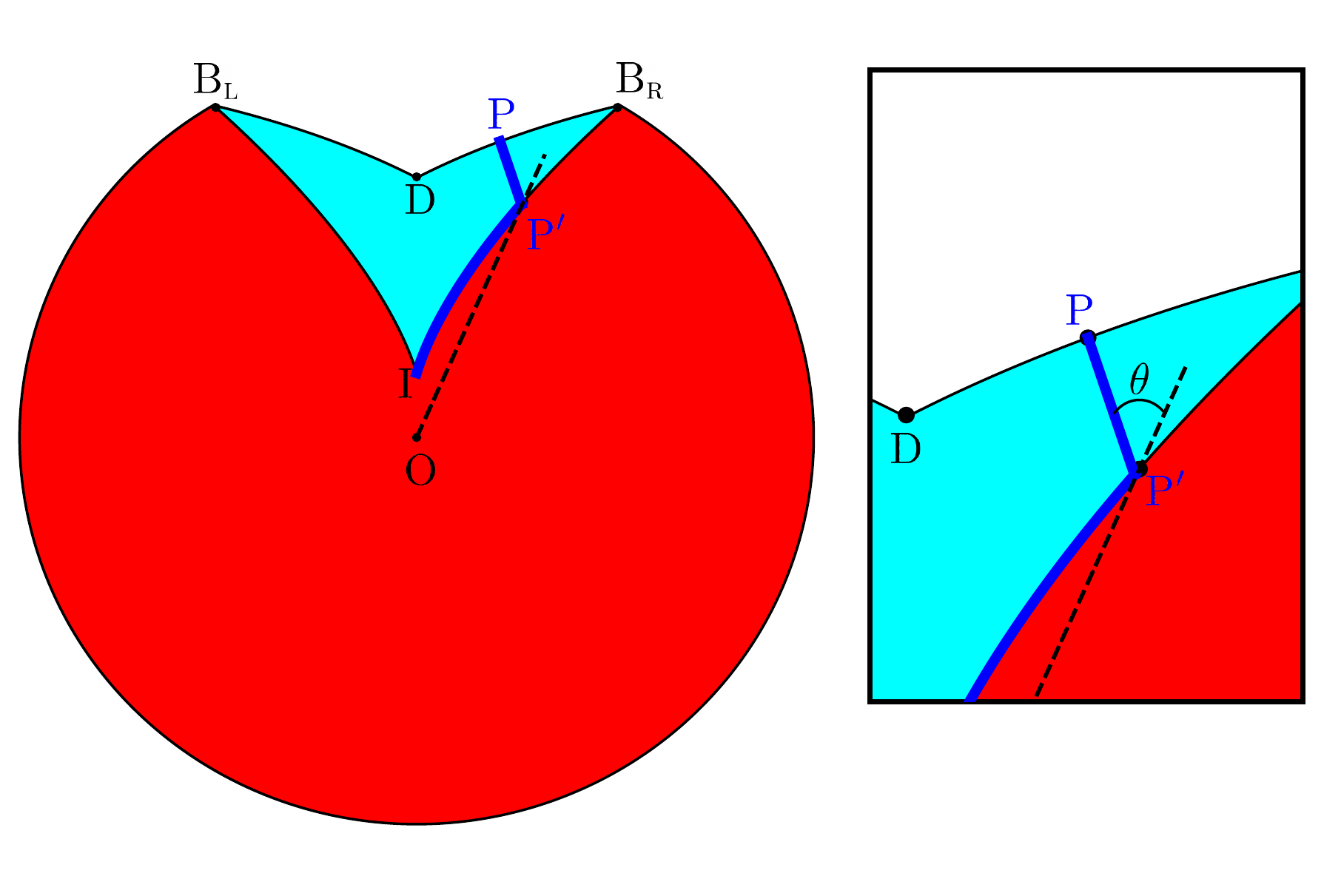}
	\centering
	\caption{Circular colony with a dented front,~$v_{\mathrm{\textsc{w}}} > v_{\mathrm{\textsc{m}}}$.  The path of the shortest time follows the sector boundary from $I$ to~$P'$ and then a straight line connecting~$P'$ and~$P$. Note that $\overline{P'P}$ and $\overline{OP'}$ always intersect at angle $\theta$. }\label{fig:g4}
\end{figure}

We found that all possible sector shapes fall into three classes. Without loss of generality, we take~$u$ to be positive by calling the mutant the strain that invades locally. The shape of the sector is then largely determined by~$v_{\mathrm{\textsc{m}}}/v_{\mathrm{\textsc{w}}}$. When this ratio is less than one, sectors have a dented shape. In the opposite case, sectors bulge outwards. The exact shape of the front of course depends on all three velocities. Overall, there are the two broad classes discussed above and a special limiting case when~$u=\sqrt{v_{\mathrm{\textsc{m}}}^2-v_{\mathrm{\textsc{w}}}^2}$ which is discussed below. In all cases, we obtained sector shapes analytically for both circular and flat initial fronts~(SI Fig.~\ref{fig:g3}, Fig.~\ref{fig:g4}). The latter are summarized in Fig.~\ref{fig:2}C and are used to test the theoretical predictions. 

The geometrical theory provides a concrete way to define local fitness advantage,~$u/v_{\mathrm{\textsc{w}}}$, and global fitness advantage,~$v_{\mathrm{\textsc{m}}}/v_{\mathrm{\textsc{w}}}-1$. These two types of fitness can take arbitrary values, even with opposite signs. The only condition is that a positive~$u$ needs to be larger than~$\sqrt{v_{\mathrm{\textsc{m}}}^2-v_{\mathrm{\textsc{w}}}^2}$ when the mutant is faster than the wildtype. This constraint arises because, for large~$v_{\mathrm{\textsc{m}}}/v_{\mathrm{\textsc{w}}}$, the gaining of new territory due to the large global fitness advantage outpaces the gain in the new territory due to a smaller local fitness advantage. The constraint on~$u$ is not relevant to dented fronts, so we relegate this discussion to the SI~(Fig.~\ref{fig:g1}).

\subsection*{Experimental test of the geometric theory}
How can we test whether the theory of sector geometry described above indeed applies to our experiments? The theory utilizes three velocities~$v\w$,~$v\m$, and~$u$ to predict the shape of the sector boundary and the sector front. The absolute values of the velocities determine how quickly the colony grows overall and its shape depends only on two dimensionless parameters:~$v\m/v\w$~and~$u/v\w$. The first parameter can be obtained from the direct measurements of expansion velocities in monocultures. The second parameter can be inferred by fitting the shape of the sector boundary to the theory. This leaves the shape of the sector front as an independent measurement that can be compared to the theoretical prediction. 

\begin{figure}[htp]
    \centering
    \includegraphics[width=.8\linewidth]{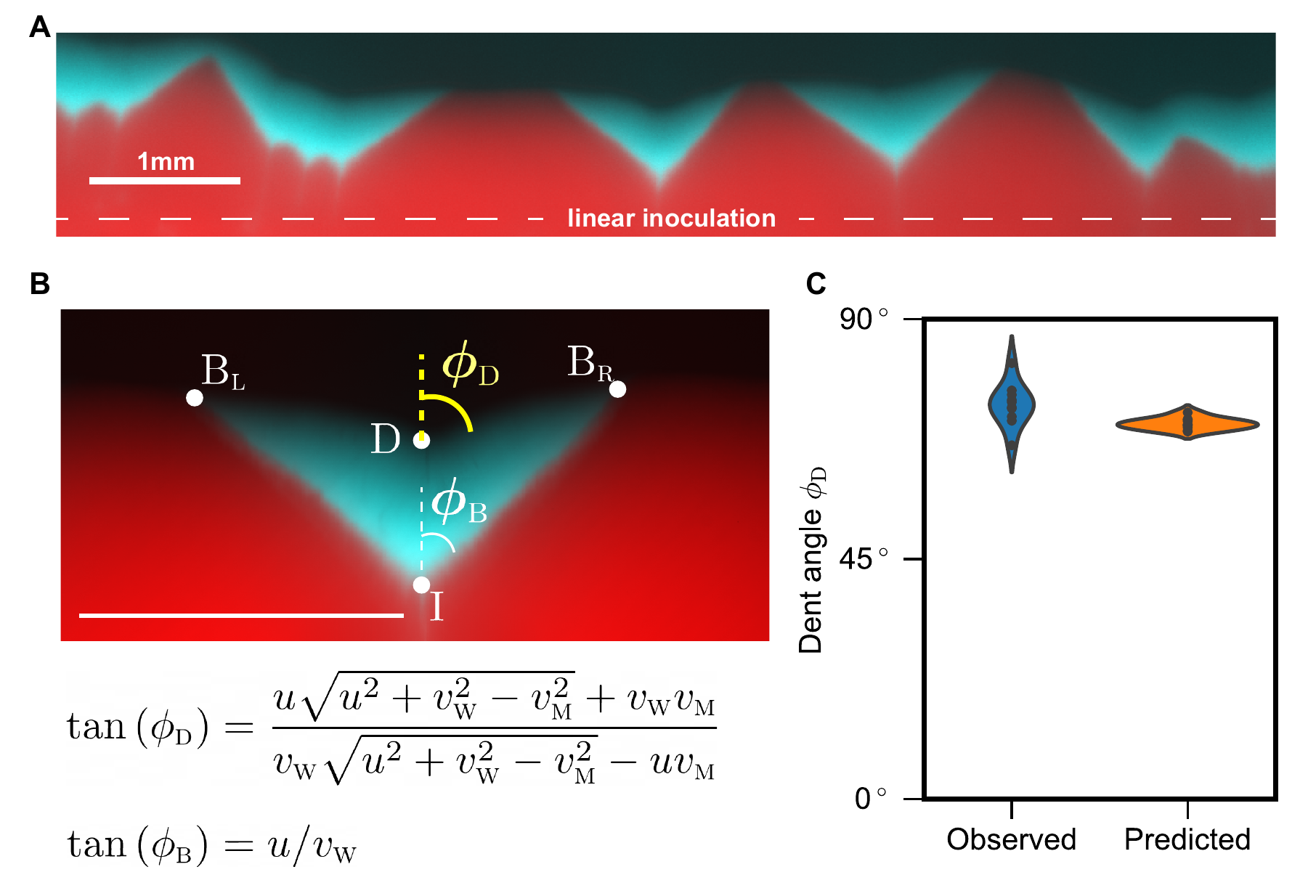}
    \caption{Empirical test of predicted sector shapes. (A)~We used linear inoculations with low density and low fraction of the mutant and grew the colonies for 48~hours. (B)~Top: Zoom-in image of one of the sectors. The shape of mutant sector is quantified by two opening angles: one between the two sector boundaries~$2 \phi_\mathrm{\textsc{b}}$ and one between the two parts of the expansion front that meet at the dent~$2 \phi_\mathrm{\textsc{d}}$. Bottom: The theory predicts $\phi_\mathrm{\textsc{b}}$ and $ \phi_\mathrm{\textsc{d}}$ as functions of the three velocities: $v\w$, $v\m$, and $u$. We used~$\phi_\mathrm{\textsc{b}}$ to determine $u / v\w$ and predict $\phi_\mathrm{\textsc{d}}$; $v\m/v\w$ is measured from monoculture expansions. (C)~The observed and predicted values of $\phi_\mathrm{\textsc{d}}$ are very close to each other.}
    \label{fig:3}
\end{figure}

The linear expansion geometry greatly simplifies all the steps involved in testing the theory because the shapes of both the sector boundary and the dent are determined by their opening angles. Qualitative agreement with this theoretical prediction is quite clear from the experimental images~(Fig.~\ref{fig:3}A), which indeed show that mutant sectors are bounded by straight lines on all sides. The opening angle of the sector boundary determines~$u/v\w$ and the opening angle of the dent serves as a testable prediction~(Fig.~\ref{fig:3}B).

Our experiments proceeded as follows. We first measured expansion velocities in monocultures by tracking the colony radius as a function of time; see Fig.~\ref{fig:1}B. Then, the data on sector shapes were collected from plates inoculated along a straight line with a low-density ($10^{-3}$ $\mathrm{OD_{600}}$) 99:1 mixture of the wildtype and the mutant. After two days of growth, five well-isolated sectors were analyzed to determine~$\phi_\mathrm{\textsc{b}}$ and~$\phi_\mathrm{\textsc{d}}$~(see Methods). Since each side of the angle can be used, we effectively obtained ten measurements. Figure~\ref{fig:3}C shows that observed~$\phi_\mathrm{\textsc{d}}$ is~$73.93^\circ$ (SD=$3.81^\circ$,~SEM=$1.21^\circ$,~n=10). Predicted~$\phi_\mathrm{\textsc{d}}$ is~$70.39^\circ$ (SD=$1.02^\circ$,~SEM=$0.32^\circ$,~n=10). This is an excellent agreement given other sources of variability in our experiment including variations in velocity between replicates and potential systematic errors in fitting sector shapes. Thus the geometric theory not only provides an explanation of the novel sector shape, but also describes it quantitatively.

Another experimental verification of our theory comes from Ref.~\cite{korolev:sectors} that studied sector shapes in yeast colonies. Instead of dents, their strains produced circular bulges (the special case with $v\m = \sqrt{v\w^2 + u^2}$). For this special case, our results fully agree with both their theoretical and experimental findings (see Eq.~12, Table~1, and Figure~S8 in Ref.~\cite{korolev:sectors}). As far as we know, the intermediate case of composite bulge (see Fig.~\ref{fig:2}C) has not been observed yet. Perhaps engineered strains with a tunable tradeoff between local and global fitness would enable the observation of all sector shapes in a single system.
 
\subsection*{Concrete mechanisms of fitness tradeoff}
The geometric theory integrates local and global competition and quantitatively predicts the shape of mutant sector in our experiment. Yet, the theory does not provide a tangible mechanism behind the takeover by a slower expander. To show that dented fronts emerge readily under different ecological scenarios we used the flexible framework of reaction-diffusion models, which are also known as generalized Fisher-Kolmogorov equations~\cite{fisher1937wave, kolmogorov1937study, murray2002mathmatical}. A general model can be written as:

\begin{equation}
\begin{aligned}
& \partial_t n\w =  \left(\nabla^2 \left( D\w n\w \right) + r\w n\w\right) (1-n\w-n\m),\\
& \partial_t n\m = \left(\nabla^2 \left( D\m n\m \right) + r\m n\m\right) (1-n\w-n\m).
\end{aligned}
\label{rd}
\end{equation}

Here,~$n\w$ and~$n\m$ are the population densities of the wildtype and the mutant normalized by the shared carrying capacity; $D\w$, $r\w$ and $D\m$, $r\m$ are their respective dispersal and per capita growth rates. 

The factor of $(1-n\w-n\m)$ ensures that there is no growth or movement behind the front. In the growth term, this is a standard assumption that ensures finite carrying capacity~\cite{murray2002mathmatical}. In the dispersal term, the factor of $(1-n\w-n\m)$ has rarely been studied in mathematical biology because it is specific to microbial range expansions, where there is no movement behind the front~\cite{korolev:sectors, korolevAmericanNaturalist, muller2014genetic, momeni2013strong}. In the SI, we demonstrate that dented fronts also occur with standard density-independent dispersal and therefore could be relevant for range expansions of macroscopic organisms~(SI appendix~\ref{sec:const}). 

The non-spatial limit of Eq.~\ref{rd} is obtained by dropping the dispersal term. This limit is analyzed in the appendix~\ref{sec:ode}. As population grows from any initial condition, the relative abundance of the faster grower increases until the total population density reaches the carrying capacity. At this point there is no further change in $n\w$ and $n\m$. This neutral coexistence between the two strains ensures that the population is frozen behind the front and the competition unfolds only at expansion frontier. 

\begin{figure}
\centering
\includegraphics[width=\textwidth]{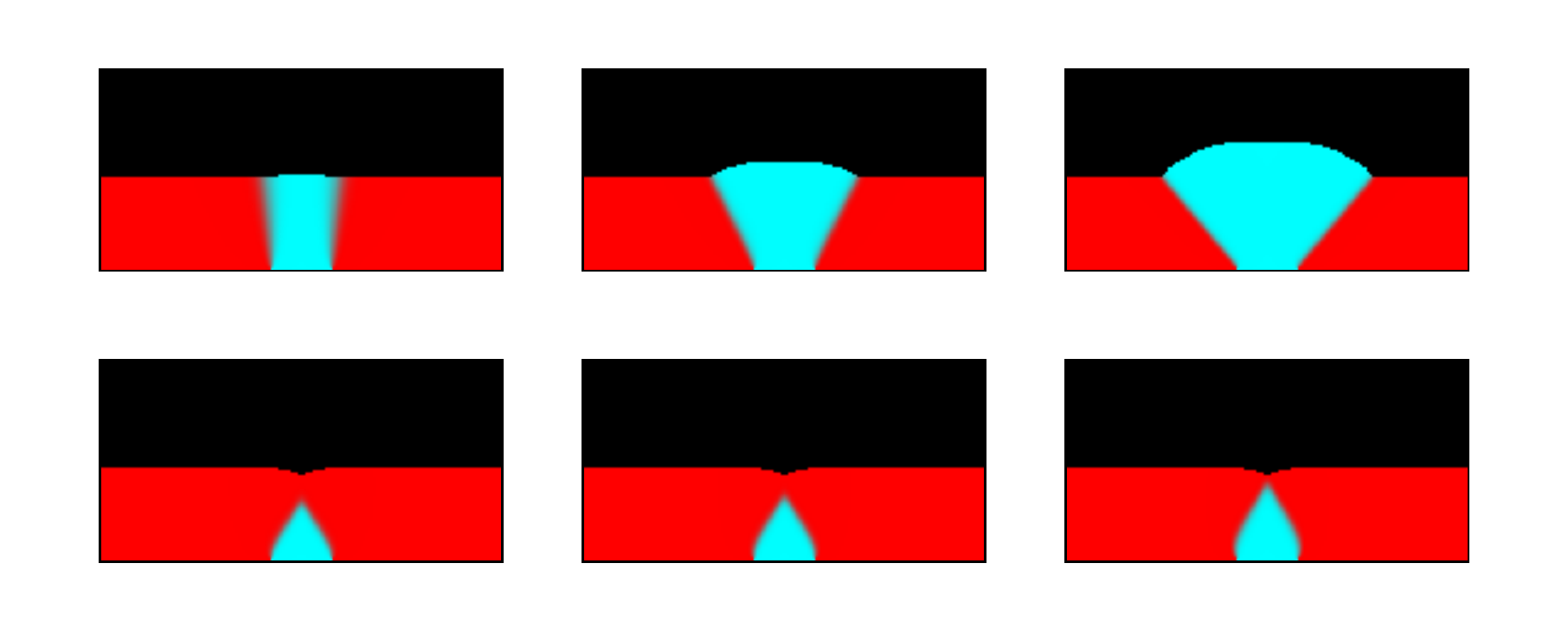}
\caption{No dented fronts occur in simulations with density-independent growth and dispersal. In each column, the growth advantage $r\m/r\w - 1$ is the same~(Left: $0.04$, Middle: $0.36$, Right:$1$). Simulations in top row have $D\w = D\m$, so that the ratio of the expansion velocities varies with the growth rates~($v\m = v\w\sqrt{r\m/r\w}$). For the bottom row, we used $D\m = \frac{0.64 r\w}{r\m} D\w$ so that $v\m = 0.8 v\w $. We observed no expanding mutant sectors when its expansion velocity was less than that of the wildtype. }\label{fig:si6}
\end{figure}

The simplest spatial model takes all growth and dispersal rates to be independent of population density. It is then easy to show that there is no difference between local and global fitness; see Fig.~\ref{fig:si6} and Ref.~\cite{korolev:sectors}. Most of the previous work focused on this special case of so-called ``pulled'' waves~\cite{birzu2018fluctuations} and thus could not observe the takeover by the slower expander.

Many organisms, however, exhibit some density dependence in their growth or dispersal dynamics~\cite{levin1969dependence, courchamp1999inverse, matthysen2005density, kearns2010field, peischl2020evolution}, which can lead to a tradeoff between local and global fitness. One commonly-studied case is found in the interaction between cooperators and cheaters~\cite{nadell2010emergence,korolev2013fate,yan2017extracellular,cremer2019cooperation}. To model this ecological scenario, we take

\begin{equation}
\begin{aligned}
&   D\w = D\m = D,\\
&   r\w = r\left(1 - \alpha \frac{n\m}{n\w + n\m} \right), \quad r\m = r\left(1 - s + \alpha \frac{n\w}{n\w + n\m} \right).\\
\end{aligned}
\label{rd1}
\end{equation}

The benefit of cooperation is specified by~$s$, which is the difference in the growth rate of cooperators and cheaters when grown in isolation. The benefit of cheating is controlled by~$\alpha$; the growth rate of cheaters increases by up to~$\alpha$ provided cooperators are locally abundant. For simplicity, we chose a symmetric linear dependence of the growth rates on the mutant frequency and assumed that the diffusion constants are equal. 

\begin{figure}[htp]
    \centering
    \includegraphics[width=.8\linewidth]{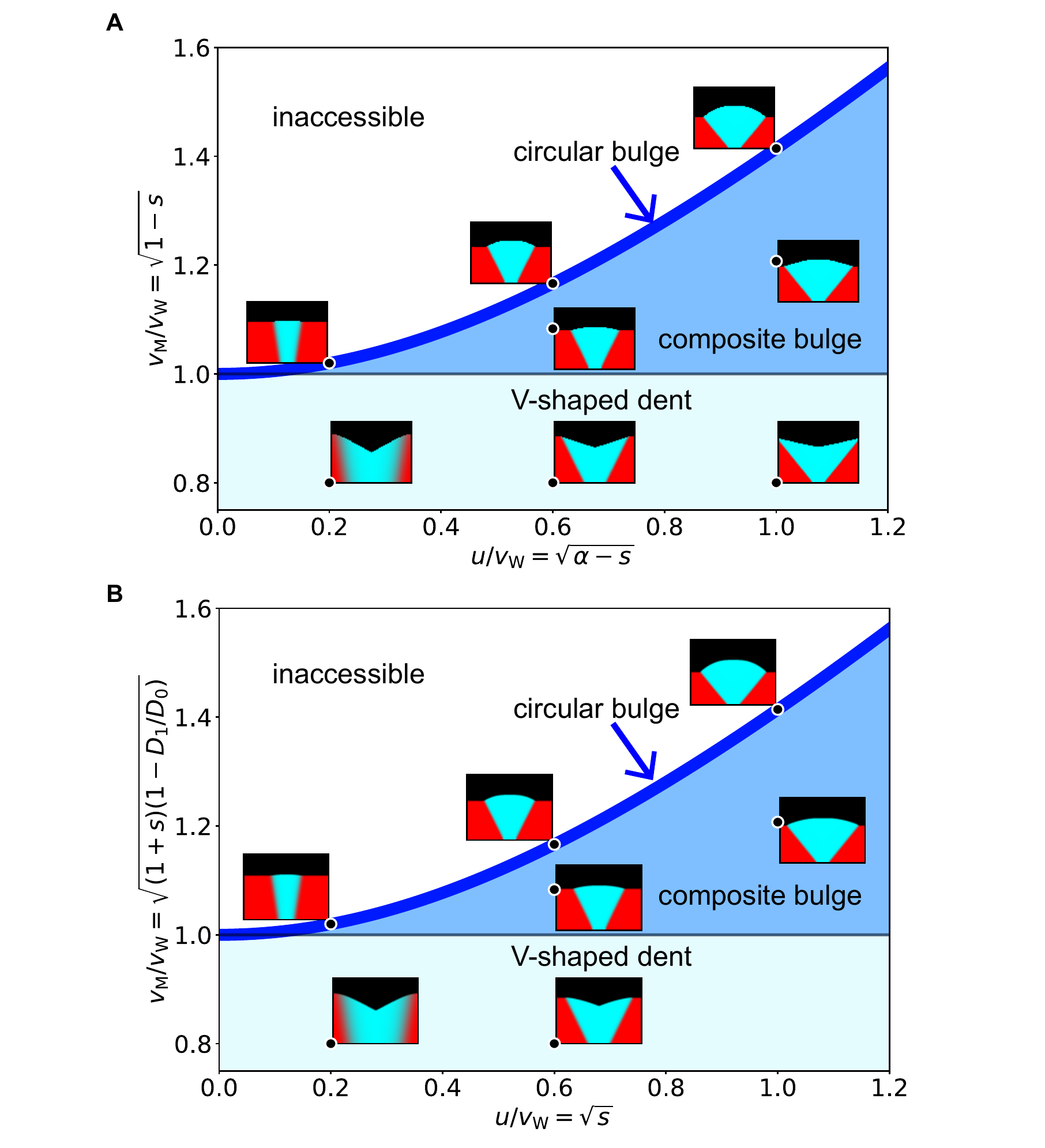}
    \caption{Sector shapes from microscopic simulations recapitulate phase diagram from the geometric theory. (A)~Simulation of cooperator-cheater model~(Eq.~\ref{rd1}) is compared with the geometric theory. By varying $s$~(benefit from cooperation) and $\alpha$~(strength of cheating), we explored sector shapes for different values of $v\m/v\w$ and $u/v\w$. The locations of various sector shapes match the predictions of the geometric theory. In particular, V-shaped dents are observed when a cheater expands more slowly than a cooperator~($s>0$), but has a sufficiently large advantage from cheating~($\alpha>s$). (B)~Simulations of growth-dispersal tradeoff model~(Eq.~\ref{rd2}) also agree with the geometric theory. Different sector shapes were obtained by varying the the growth advantage~$s$ and and the dispersal disadvantage~$D_1$. See Methods for simulation parameters.}
    \label{fig:4}
\end{figure}

Numerical simulations of this model reproduced a V-shaped dented front~(Fig.~\ref{fig:4}A). The dents flattened when there was no benefit to cooperate~($s=0$) and were replaced by bulges when cooperators grew slower than cheaters~($s<0$). We were also able to test whether these transitions in sector shape matched the predictions of the geometric theory. For this comparison between the theory and simulations, we need a mapping between the microscopic parameters of the model and the three velocities that enter our geometric theory. Fortunately,{ in this model,} all three velocities can be calculated analytically: $v\w = 2\sqrt{r D(1+s)}$, $v\m = 2\sqrt{r D}$, and $u = \sqrt{(\alpha-s)r D}$. Therefore, we could overlay individual simulations on the phase diagram predicted by the geometric theory. The result, shown in Fig.~\ref{fig:4}A, shows the expected agreement and provides further validation for the geometric theory.

The geometric description is generic and should transcend the specifics of the cooperator-cheater model discussed above. To further illustrate that different ecological interactions can produce identical spatial patterns, we simulated a completely different mechanism for the tradeoff between local and global fitness. This time, we assumed that the wildtype loses the local competition because it grows slower than the mutant, but this slower growth is more than compensated by a much higher dispersal rate. This growth-dispersal tradeoff may be common in nature~\cite{tilman1994competition,bonte2012costs,fraebel2017environment,gude2020bacterial}, and is captured by the following set of parameters:

\begin{equation}
\begin{aligned}
&   D\w = D\m = D_0 - D_1 \frac{n\m}{n\w + n\m}  ,\\
&   r\w = r, \quad r\m = r\left(1 + s \right).\\
\end{aligned}
\label{rd2}
\end{equation}

\begin{figure}
\centering
\includegraphics[width=\textwidth]{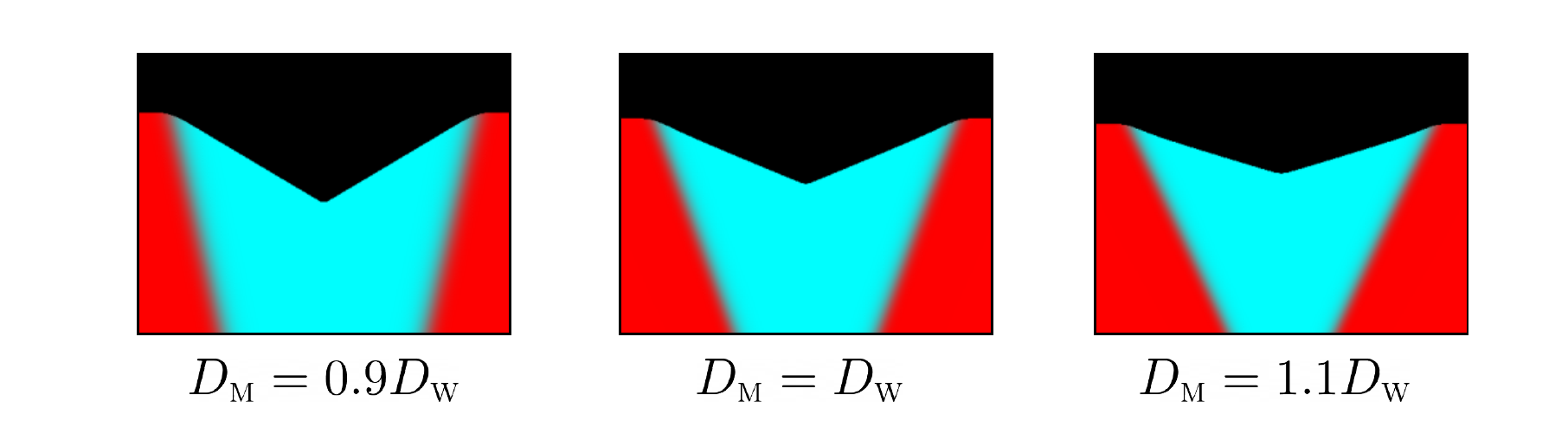}
\caption{Dented fronts occur in simulations with $D\m \neq D\w$. We used a variation of cheater-cooperator model (Eq.~\ref{rd1}) in which dispersal of wildtype and mutant is no longer identical. In all cases of $D\m=0.9 D\w$, $D\m= D\w$, and $D\m=1.1 D\w$ mutant developed a dented front with only quantitative changes in sector shapes. These simulations used parameters $s=0.4$ and $\alpha=0.6$. }\label{fig:si7}
\end{figure}

Here, the growth rates are density-independent, but the dispersal rates change with the local community composition. We chose~$D\w=D\m$ to reflect the collective nature of movement in colonies of non-motile microbes~\cite{farrell2013mechanically,warren2019spatiotemporal}, which are pushed outward by mechanical stressed generated by all cells behind the front. In addition, this simplifying assumption enables us to calculate the velocities analytically and construct a quantitative phase diagram similar to Fig.~\ref{fig:4}A. In the SI, we show that dented front can also be observed in models with~$D\w \neq D\m$ (Fig.~\ref{fig:si7}).

Our simulations again exhibited dented fronts and all shape transitions in full agreement with the geometric model~(Fig.~\ref{fig:4}B). Thus, the geometric description is universal, i.e. a wide set of growth-dispersal dynamics converges to it. This universality, however, makes it impossible to determine the specifics of ecological interactions from spatial patterns alone. In other words, the observation of a dented front indicates the existence of a tradeoff between local and global fitness, but does not hint at any specific mechanism that is responsible for this tradeoff. For example, both models (Eq.~\ref{rd1} and Eq.~\ref{rd2}) produce identical sector shapes (Fig.~\ref{fig:4}) and both would provide a perfect fit to our experimental data. Indeed, each model has four parameters, which is more than sufficient to specify the three velocities that control all aspects of spatial patterns. Such fits of course would not provide a meaningful insight into the mechanism. To determine the mechanism, one would have to perform a different kind of experiments that could probe population dynamics on the spatial scale of local competition. 

\section*{Discussion}

This study used a simple and well-controlled laboratory microcosm to elucidate the factors that influence spatial competition. We found a stark contradiction to the intuitive expectation that the faster runner wins the race~\cite{deforet2019evolution}. A mutant that expanded more slowly on its own nevertheless took over the expansion front when inoculated with the wildtype. This spatial takeover accompanied V-shaped sectors, which are a characteristic signature of the mismatch between local and global competition. To explain these observations, we developed a theory that integrates local and global competition and predicts all possible sector shapes. We then confirmed the validity of the theory using both further experiments and simulations.

Our experimental results unequivocally demonstrate that a slow expander can win with and without sectoring. Under low genetic drift conditions, the slow expander took over the front uniformly across the colony. This outcome can be described by one-dimensional models because the competition occurs primarily along the radial direction. In contrast, stronger genetic drift resulted in sector formation and produced fully two-dimensional growth dynamics. Even under these less favorable conditions, the slower mutant still outcompeted the wildtype. 

Previously, slower expanders were found to be successful only in one-dimensional models~\cite{travis2002dispersal,korolev2015evolution,datta2013range}, and only buldged sectors of faster expanders were reported for two-dimensional growth~\cite{korolev:sectors}. The latter was true even when there was a tradeoff between local and global fitness~\cite{van2013spatial}, presumably because local fitness advantage was not sufficiently large. Our experiments not only confirm the predictions of one-dimensional models, but also expand the set of conditions under which the unusual takeover by a slower mutant can be observed. In fact, the slower expanders could be successful in many settings not only because the theory and simulations strongly support this claim, but also because we relied on evolved mutants from natural isolates rather than genetic engineering to obtain the strains. 

The observation of dented fronts clearly shows that the existing theoretical understanding of sector growth is incomplete. Previously, it was assumed that the spatial pattern depends only on the ratio of the mutant and wildtype velocities~\cite{korolev:sectors}. This simple picture holds when the fast expander also has a moderate advantage in local competition. More generally, however, we found that the outcome of the competition also depends on the velocity~$u$ with which one of the strains invades locally. The sector shapes are completely determined by the three velocities~($v\m$,~$v\w$,~$u$) and can be used to make quantitative inferences from experimental data. Nevertheless, the main contribution of our theory is its ability to integrate local and global competition and predict how large scale spatial patterns emerge from species interactions.

The geometric theory is not without limitations. This phenomenological theory cannot predict whether the fast or the slow mutant wins in a given system. To answer that question, one needs to consider a mechanistic model and derive how the invasion velocity~$u$ depends on microscopic parameters, which we have done for specific models. The universal nature of the geometric theory also precluded us from identifying the mechanism responsible for the growth dynamics observed in our experiments. We left this fascinating question for future works, and instead, focused on several common tradeoffs between local and global fitness. The simulations of these tradeoffs not only confirmed the validity of the geometric theory, but further highlighted that slower expanders could establish by a wide range of mechanisms.

The geometric theory also relies on a few technical assumptions such as constant expansion velocities, negligible stochasticity, and the absence of long-range interaction due to chemotaxis or nutrient depletion. Relaxing these assumptions could lead to certain quantitative changes in sector shapes, but the existence of dented fronts or the possibility of a takeover by a slower expander should not be affected. 

Our work opens many directions for further investigation. We clearly showed that the expansion velocity cannot be the sole determinant of the spatial competition. Therefore, it will be important to examine how local interactions influence the eco-evolutionary dynamics during range expansions. Such future work would bring about a more detailed description of ecological and biophysical processes in growing populations. It would also greatly enhance our understanding of the tradeoffs among different life-history traits and shed light on the incredible diversity of successful strategies to navigate spatial environments~\cite{tilman1994competition,bonte2012costs,fraebel2017environment,gude2020bacterial}. The geometric theory developed here provides a convenient way to integrate these various aspects of population dynamics. It abstracts the main features of spatial growth and should facilitate the analysis of both experiments and simulations.

\section*{Material and Method}
\subsection*{Strains}
\begin{figure}[ht!]
\centering
\includegraphics[width=\textwidth]{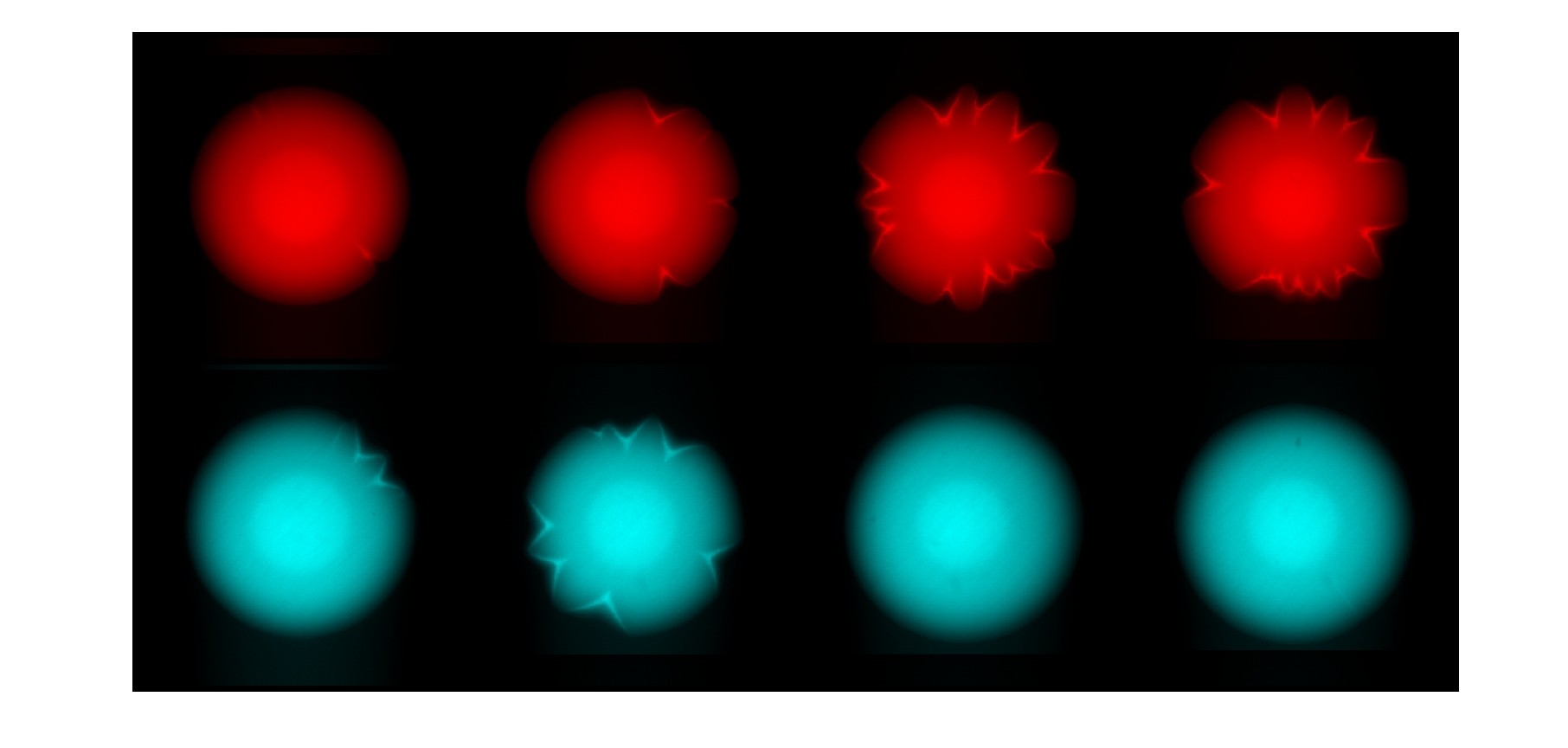}
\caption{Emergence of dents in wildtype colonies was reproducible. Wildtype colonies were grown for 48 hours. Top: wildtype strains constitutively expressing mScarlet-I. Bottom: wildtype strains constitutively expressing mTurquois-2.  }\label{fig:si1}
\end{figure}

Wildtype \textit{Raoultella planticola} strains were isolated from a soil sample (MIT Killian Court, Cambridge, MA)~\cite{kehe2019massively} and were tagged with two different fluorescent proteins mScarlet-I (red) and mTurquois2 (cyan) by insertion of plasmids pMRE145 and pMRE141 respectively~\cite{schlechter2018chromatic}.
As we grew wildtype colonies on agar plates, they reproducibly developed dents after several days as shown in Fig.~\ref{fig:1}A and Fig.~\ref{fig:si1}. We sampled the cells from either inside the dent or on the smooth edge using inoculation loops, streaked on small plates, and grown in $30^{\circ}\mathrm{C}$ for two days. Then we sampled single colonies, grew them overnight in LB growth media, and stored as a $-80^{\circ}\mathrm{C}$ glycerol stock. 
\subsection*{Growth media preparation}
We prepared hard agar plates with 1X Luria-Bertani media (LB, 2.5\% w/v; BD Biosciences-US) and 1.5\% w/v of agar (BD Bioscience-US). We also added 1X Chloramphenicol (Cm, 15mg/L, prepared from 1000X solution) for constitutive expression of fluorescence. For each agar plate, 4mL of media was pipetted into a petri dish~(60X15mm, sterile, with vents; Greiner Bio-one), and was cooled overnight (15 hours) before inoculation. 
\subsection*{Expansion experiment}
For each strain, $-80^{\circ}\mathrm{C}$ glycerol stock was streaked on a separate plate and grown for 2 days. Then a colony from each strain was picked up and put into a 50mL Falcon Tube filled with $5\;$mL of liquid media (1X LB and 1X Cm). Bacterial cultures were grown overnight at $30^{\circ}\mathrm{C}$ under constant shaking 1350 rpm (on Titramax shakers; Heidolph). We then diluted and mixed the cultures to desired total density and mutant fraction, measured in optical density~($\mathrm{OD_{600}}$) using a Varioskan Flash (Thermo Fisher Scientific) plate reader.
For circular expansions, we gently placed a droplet of $1.5\;\mu$L inoculant at the center of an agar plate. For linear expansions, we dipped a long edge of a sterile cover glass~(24X50mm; VWR) gently into the culture and touched the agar plate with the edge. After inoculation, each colony was grown at~$30^{\circ}\mathrm{C}$ for 48 hours.
\subsection*{Imaging}
At fixed times after inoculation, each plate was put on a stage of Nikon Eclipse Ti inverted light microscope system. 10X magnification was used for whole-colony images, and 40X magnification was used for single sector images. Fluorescent images were taken using Chroma filter sets ET-dsRed~(49005) and ET-CFP~(49001) and a Pixis 1024 CCD camera. \\
We used scikit-image~\cite{scikit-image} for image processing in Python. Images from different fluorescent channels were integrated after background subtraction and normalization by respective maximum intensity. The sector boundaries were identified as the furthest points from inoculation plane where both strains' FL intensities were above respective thresholds.
The codes for image analysis will be available via GitHub~(\url{https://github.com/lachesis2520/dented_front_public.git}) upon publication.
\subsection*{Numerical simulation}
Numerical simulations were performed by solving the corresponding partial differential equations on a square grid using a forward-in-time finite difference scheme that is second order accurate in space and first order accurate in time~\cite{press2007numerical}. Python codes will be available via GitHub~(\url{https://github.com/lachesis2520/dented_front_public.git}) upon publication.\\
For cooperator-cheater model simulation, we used the following set of values for parameters $(s, \alpha)$: $(-0.04, 0)$, $(-0.36, 0)$, $(-1, 0)$, $(-0.173, 0.187)$, $(-0.457, 0.543)$, $(0.36, 0.4)$, and~$(0.36, 0.72)$.\\
For growth-dispersal tradeoff model simulation, we used $(s, D_1)$ of $(0.04, 0)$, $(0.36, 0)$, $(1, 0)$, $(0.36, 0.147)$, $(1, 0.271)$, $(0.04, 0.385)$, and~$(0.36, 0.529)$.

\section*{Acknowledgements}
We thank all members of the J.G. laboratory for helpful discussions. Anthony Ortiz provided the wildtype~\textit{R.~planticola} strain and Daniel R. Amor helped with preliminary work for this study. This work was supported by NIH~(R01-GM102311) and the Sloan Foundation~(G-2021-16758) to J.G. and by the Simons Foundation Grant \#409704, Cottrell Scholar Award \#24010, and by NIGMS grant \#1R01GM138530-01 to K.S.K. The authors also acknowledge the MIT SuperCloud and Lincoln Laboratory Supercomputing Center for providing computational resources.

\clearpage
\beginsupplement

\title{\Large Supplemental Information}
\date{}
\maketitle

\tableofcontents
\clearpage
\section*{Supplementary figures}
\clearpage
\clearpage
\clearpage
\clearpage
\clearpage
\clearpage
\clearpage

\section{Geometric theory and sector shapes} \label{sec:geometric}

\subsection{Introduction}
During spatial growth in microbial colonies or other cellular aggregates, mutants appear and compete with each other. Previous studies~\cite{korolev:sectors} and common intuition suggest that advantageous mutants should form a sector that bulges out of the expansion front. In the main text, we reported experiments showing that this is not always the case. Here, we identify all possible shapes that can result from competition between two types in a growing colony.

To make progress, we make a number of approximations and work in the so-called geometrical optics limit. This limit assumes that the expansion front and the boundary between the types can be treated as thin lines. Neglecting sector and boundary widths is justified when these length scales are much smaller than the colony size. In small colonies, thin boundaries require strong genetic drift and slow motility. Furthermore, we assume that the expansion velocity of each type remains fixed. In particular, we neglect the effects of spatial variation in nutrient concentration due to protrusions of one type ahead of the other. This approximation is valid for high nutrient concentrations and when the size of the protrusions is small compared to the size of the mutant sector. 

\begin{figure}[ht!]
	\includegraphics[width=0.6\linewidth]{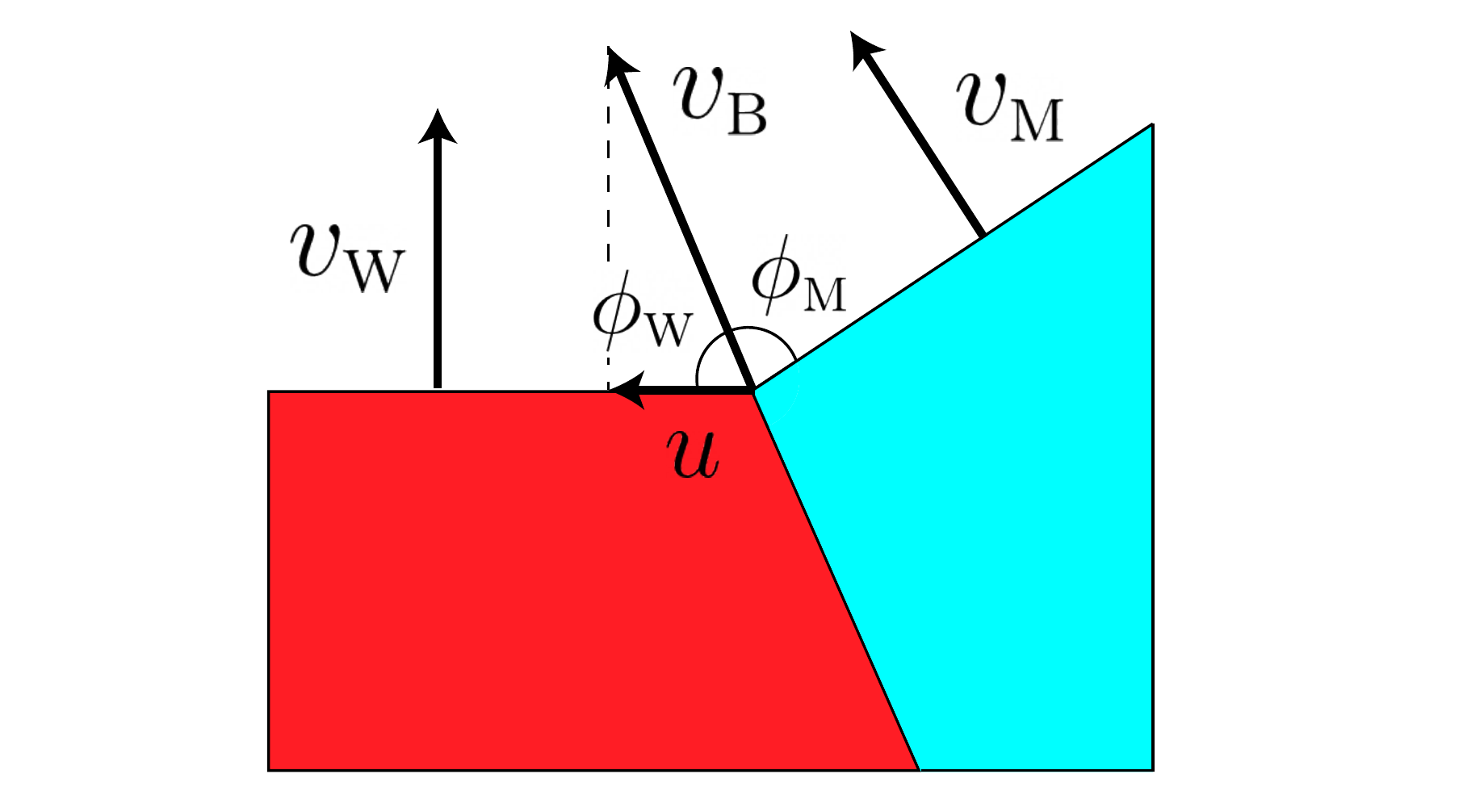}
	\centering
	\caption{ Geometry of the competition.}\label{fig:g1}
\end{figure}

In the geometric-optics limit, the competition between two types is described by three velocities: the velocity of mutant~$v_{\mathrm{\textsc{m}}}$, the velocity of wildtype~$v_{\mathrm{\textsc{w}}}$, and the velocity of the boundary~$v_{\mathrm{\textsc{b}}}$, which are shown in Fig.~\ref{fig:g1}. (Note~$v_{\mathrm{\textsc{b}}}\ne u$) Previous work~\cite{korolev:sectors} focused on the regime when~$v_{\mathrm{\textsc{b}}}$ was determined by~$v_{\mathrm{\textsc{m}}}$ and~$v_{\mathrm{\textsc{w}}}$; in contrast, we make no assumptions about the relative magnitude of these three velocities. 

In the close vicinity of the sector boundary, the two expansion fronts can be approximated as straight lines. Their position ~(Fig.~\ref{fig:g1}) is determined by requiring that the expansion along the boundary with velocity~$v_{\mathrm{\textsc{b}}}$ results in the same displacement of the fronts as moving perpendicular to them with velocities~$v_{\mathrm{\textsc{m}}}$ and~$v_{\mathrm{\textsc{w}}}$ respectively:

\begin{align}
& v_{\mathrm{\textsc{w}}} = v_{\mathrm{\textsc{b}}}\sin\phi_{\mathrm{\textsc{w}}},\\
& v_{\mathrm{\textsc{m}}} = v_{\mathrm{\textsc{b}}}\sin\phi_{\mathrm{\textsc{m}}}.
\label{joint_angles}
\end{align}

For linear inoculations, the above equations are sufficient to completely specify sector shapes because, as we show below, the expansion fronts are straight lines even away from the sector boundary. For circular initial conditions, Eqs.~(\ref{joint_angles}) provide information only about the local orientation at the sector boundary, and further calculations are necessary. One way to obtain global shape is to write down partial differential equations that specify how the position of the front changes and use Eqs.~(\ref{joint_angles}) as the boundary conditions. A much simpler approach is to use an equal time argument from Ref.~\cite{korolev:sectors}.

This method traces the ancestral lineage from each point along the front and requires that the time traveled on that lineage is equal to the current time~$t$. The location of the ancestral lineage is such that it takes the shortest time to reach the initial population starting from a given point without entering the space occupied by the other type. The details of these calculations are provided below.

Before proceeding, we note that, here and in the main text, we typically parameterize the problem with velocity~$u$ rather than~$v_{\mathrm{\textsc{b}}}$. Since~$u$ is defined as the velocity of the boundary point along the front of wildtype, we can obtain it by projecting the boundary velocity on the expansion front of the wildtype:

\begin{equation}
u=v_{\mathrm{\textsc{b}}}\cos \phi_{\mathrm{\textsc{w}}}.
\end{equation}

From this equation and Eq.~(\ref{joint_angles}), it follows that

\begin{equation}
v_{\mathrm{\textsc{b}}} = \sqrt{v_{\mathrm{\textsc{w}}}^2 + u^2}.
\label{vb}
\end{equation}

In the following, we assume that mutant takes over the front, i.e.~$u>0$. Mutants with negative~$u$ immediately become extinct at least in the deterministic model considered here.
 
Finally, we observe that Eqs.~(\ref{joint_angles}) impose constraints on the values of the three velocities. In particular, since sines are always less than one, the boundary velocity~$v_{\mathrm{\textsc{b}}}$ must be greater or equal than both~$v_{\mathrm{\textsc{m}}}$ and~$v_{\mathrm{\textsc{w}}}$. In terms of~$u$, this implies that

\begin{equation}
v_{\mathrm{\textsc{m}}}\le \sqrt{u^2+v_{\mathrm{\textsc{w}}}^2}.
\label{constraint}
\end{equation}

\subsection{Linear inoculation}
\begin{figure}[ht!]
	\includegraphics[width=0.8\linewidth]{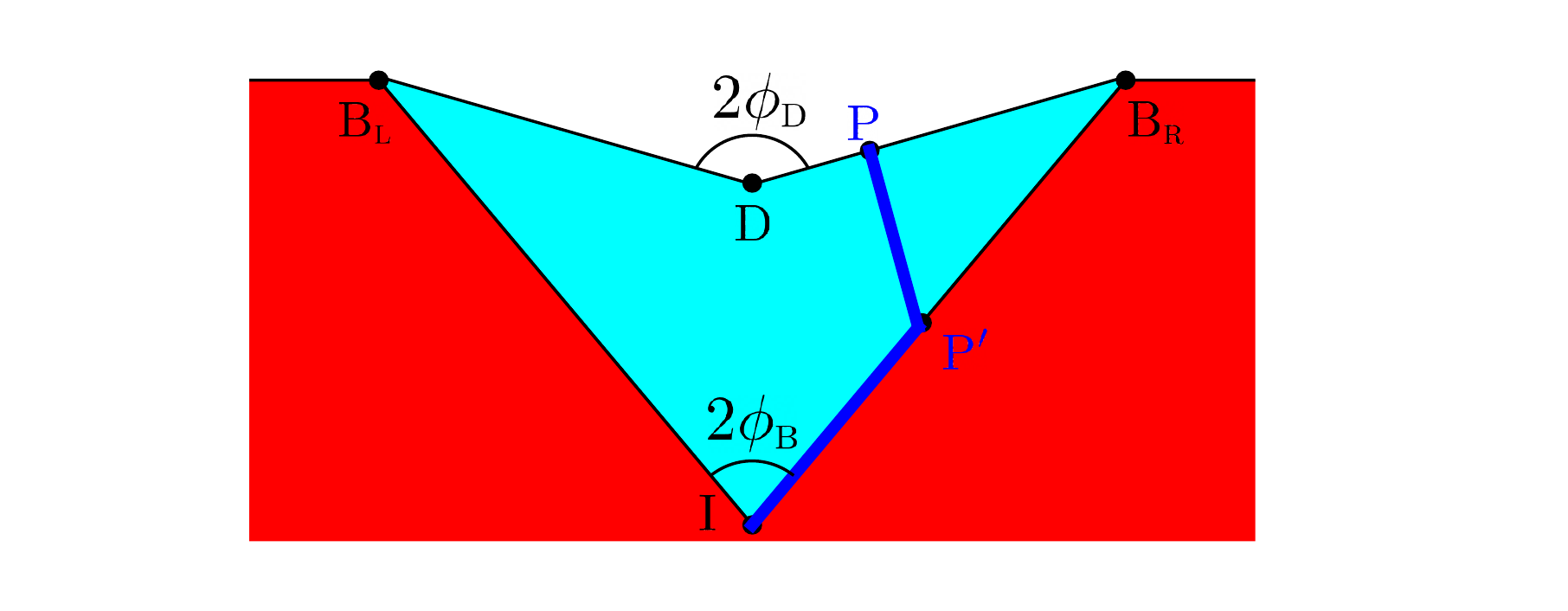}
	\centering
	\caption{ Sector shape for linear inoculation and $v_{\mathrm{\textsc{m}}} < v_{\mathrm{\textsc{w}}}$. Sectors of faster wildtype (red) and slower mutant (cyan) meet at sector boundary $\overline{IB}_\mathrm{\textsc{l}}$ and $\overline{IB}_\mathrm{\textsc{r}}$. It takes the shortest time for the mutant to go from its initial location at~$I$ to a point on the front~$P$ by first following $\overline{IP'}$ and then $\overline{P'P}$ (blue path). The resulting geometry can be characterized by two opening angles:~$2\phi_\mathrm{\textsc{b}}$ for the sector boundary and~$2\phi_\mathrm{\textsc{d}}$ for the expansion front. }\label{fig:g2}
\end{figure}
\subsubsection{Sector boundary}
Linear expansion geometry, the simplest situation to consider, allows us to explain the essence of the equal time argument. This geometry is illustrated in Fig.~\ref{fig:g2}. Initially~($t=0$), the colony front is located at~$y=0$, and expansion proceeds in the upper half-plane. Mutant is only present at a single point, which we put at~$x=0$; the rest of the front is occupied by the wildtype. 

As the expansion proceeds, the region near~$x=0$ is affected by the competition between the types. From the definition of~$u$, the extent of this region is given by~$x\in(-ut, ut)$. Regions further away are however unaffected and expand as if only wildtype is present. Thus, for~$|x|\ge ut$, the front is located at~$y=v_{\mathrm{\textsc{w}}}t$. From these considerations, we can further conclude that the sector boundary is described by~$(ut, v_{\mathrm{\textsc{w}}}t)$. Note that, below, we consider only the right side of the expansion; the left side is described by the mirror image with respect to the y-axis. Thus,

\begin{equation}
\label{phi_b}
\tan\phi_\mathrm{\textsc{b}} = \frac{u}{v_{\mathrm{\textsc{w}}}}.
\end{equation}

Note that,~$\phi_\mathrm{\textsc{b}}=\phi_{\mathrm{\textsc{w}}} - \pi/2$, which is clear from Figs.~\ref{fig:g1} and~\ref{fig:g2}.

The shape of the front for~$|x|<ut$ depends on the relative values of~$v_{\mathrm{\textsc{m}}}$, $v_{\mathrm{\textsc{w}}}$, and~$u$.

\subsubsection{$v_{\mathrm{\textsc{m}}} \le v_{\mathrm{\textsc{w}}}$}
When mutant is slower than wildtype, we find that front has a V-shaped dent with an opening angle~$2\phi_\mathrm{\textsc{d}}$ as shown in Fig.~\ref{fig:g2}. To derive this result, we take a point~$P$ on the front with yet unknown coordinates~$(x_p,y_p)$. Note that~$x_p\in(0,ut)$. Then, we should obtain the location of the ancestral lineage that connects this point to the initial location of the mutant: point~$I$. Because the ancestral lineage is located so that to minimize the travel time, it must be a union of straight lines. Indeed, it is a well-known fact from geometrical optics that light rays travel on straight lines except where the value of the refraction index changes~\cite{born2013principles}. In our case, this means that the ancestral lineages of mutant can consist of straight lines within the mutant sector and regions of the boundary. Obviously, the ancestral lineage of the mutant cannot penetrate the region occupied by the wildtype. 

The equal time argument then offers us two possibilities: a direct connection~$\overline{IP}$ and an indirect connection~$\overset{\Huge\frown}{IP'P}$ via a point~$P'$ on the sector boundary. The times to traverse these paths are

\begin{align}
&T_{PI}=|PI|/v_{\mathrm{\textsc{m}}},\\
&T_{PP'I}= |PP'|/v_{\mathrm{\textsc{m}}} + |P'I|/v_{\mathrm{\textsc{b}}}.
\end{align}

To complete the analysis, we need to choose the path with the lowest travel time and determine all locations of~$P$ for which the travel time equals~$t$. For the direct connection, it is clear that~$P$ must lie on an arc of a circle with the radius of~$v_{\mathrm{\textsc{m}}}t$ centered at~$I$. For the indirect connection, we first need to determine the location of~$P'$, which must minimize the travel time. 

Since~$P'$ lies on the sector boundary its coordinates are given by~$(ut', v_{\mathrm{\textsc{w}}}t')$ with an unknown~$t'$. The travel time is then given by

\begin{equation}
\label{linear_T_1}
T_{PP'I} = \frac{\sqrt{(x_p-ut')^2 + (y_p-v_{\mathrm{\textsc{w}}}t')^2}}{v_{\mathrm{\textsc{m}}}} + \frac{\sqrt{u^2+v_{\mathrm{\textsc{w}}}^2}t'}{v_b}. 
\end{equation}

Upon minimizing~$T_{PP'I}$ with respect to~$t'$, we find that

\begin{equation}
t' = \frac{ u\sqrt{u^2 + v_{\mathrm{\textsc{w}}}^2 - v_{\mathrm{\textsc{m}}}^2} + v_{\mathrm{\textsc{m}}} v_{\mathrm{\textsc{w}}}  }{(u^2+ v_{\mathrm{\textsc{w}}}^2) \sqrt{u^2 + v_{\mathrm{\textsc{w}}}^2 - v_{\mathrm{\textsc{m}}}^2}    }\left( x_p + \frac{u v_{\mathrm{\textsc{w}}} - v_{\mathrm{\textsc{m}}} \sqrt{u^2 + v_{\mathrm{\textsc{w}}}^2 - v_{\mathrm{\textsc{m}}}^2}}{ u^2 - v_{\mathrm{\textsc{m}}}^2 } y_p \right),
\end{equation}

and the travel time equals

\begin{equation}
T_{PP'I} = \frac{ (u v_{\mathrm{\textsc{m}}} - v_{\mathrm{\textsc{w}}} \sqrt{u^2 + v_{\mathrm{\textsc{w}}}^2 - v_{\mathrm{\textsc{m}}}^2})x_p + (v_{\mathrm{\textsc{m}}} v_{\mathrm{\textsc{w}}} + u \sqrt{u^2 + v_{\mathrm{\textsc{w}}}^2 - v_{\mathrm{\textsc{m}}}^2})y_p }{ (u^2 + v_{\mathrm{\textsc{w}}}^2) v_{\mathrm{\textsc{m}}} } ,
\label{linear_T_2}
\end{equation}

which is smaller than~$T_{PI}$ as long as~$v_{\mathrm{\textsc{m}}}<v_{\mathrm{\textsc{w}}}$. Thus, the ancestral lineages takes an indirect path that first connects point~$P$ to the sector boundary and then follows the sector boundary until~$I$. The shape of the front is determined by setting~$T_{PP'I}$ from Eq.~(\ref{linear_T_2}) equal to~$t$. This results in a segment of a straight line, and a straightforward calculation shows that

\begin{equation}
    \phi_{\mathrm{\textsc{d}}} = \arctan\left( \frac{ u\sqrt{v_{\mathrm{\textsc{w}}}^2 + u^2 - v_{\mathrm{\textsc{m}}}^2} + v_{\mathrm{\textsc{m}}} v_{\mathrm{\textsc{w}}}} {v_{\mathrm{\textsc{w}}} \sqrt{v_{\mathrm{\textsc{w}}}^2 + u^2 - v_{\mathrm{\textsc{m}}}^2} - u v_{\mathrm{\textsc{m}}} } \right).
\end{equation} 

Because the front and the sector boundaries are straight, the result above also directly follows from Eqs.~(\ref{fig:g1}). Indeed, a simple geometric argument shows that~$\phi_{\mathrm{\textsc{d}}} = \phi_{\mathrm{\textsc{m}}} + \phi_{\mathrm{\textsc{w}}} -\pi/2$.

Note that, for~$v_{\mathrm{\textsc{m}}}=v_{\mathrm{\textsc{w}}}$, the angle~$\phi_{\mathrm{\textsc{d}}}=\pi/2$ and the whole front is flat as it should if the expansion rates of the strains are identical.

\subsubsection{$v_{\mathrm{\textsc{m}}} = \sqrt{v_{\mathrm{\textsc{w}}}^2+u^2}$}
In the limiting case of maximal allowed~$v_{\mathrm{\textsc{m}}}$, the shape of the sector is also simple and immediately follows from the calculations above. Now, as we compare the two alternative paths, we find that~$T_{PI}$ is always smaller than~$T_{PP'I}$. Thus, the shape of the sector is an arc of a circle of radius~$v_{\mathrm{\textsc{m}}}t$ around~$I$ that connects to the flat front of the wild type at the sector boundary. 

Previous work that used the equal time argument to describe competition in microbial colonies only considered~$v_{\mathrm{\textsc{m}}} = \sqrt{v_{\mathrm{\textsc{w}}}^2+u^2}$ and missed other possible front shapes~\cite{korolev:sectors}. While it might appear that~$v_{\mathrm{\textsc{m}}} = \sqrt{v_{\mathrm{\textsc{w}}}^2+u^2}$ is a very special case, this relationship between the velocities holds across a wide set of conditions. Specifically,~$v_{\mathrm{\textsc{m}}} = \sqrt{v_{\mathrm{\textsc{w}}}^2+u^2}$ whenever local competition between the types is not strong enough to alter the priority effects due to different expansion velocities. 

\subsubsection{$v_{\mathrm{\textsc{w}}} < v_{\mathrm{\textsc{m}}} < \sqrt{v_{\mathrm{\textsc{w}}}^2+u^2}$}
\begin{figure}[ht!]
	\includegraphics[width=1.\linewidth]{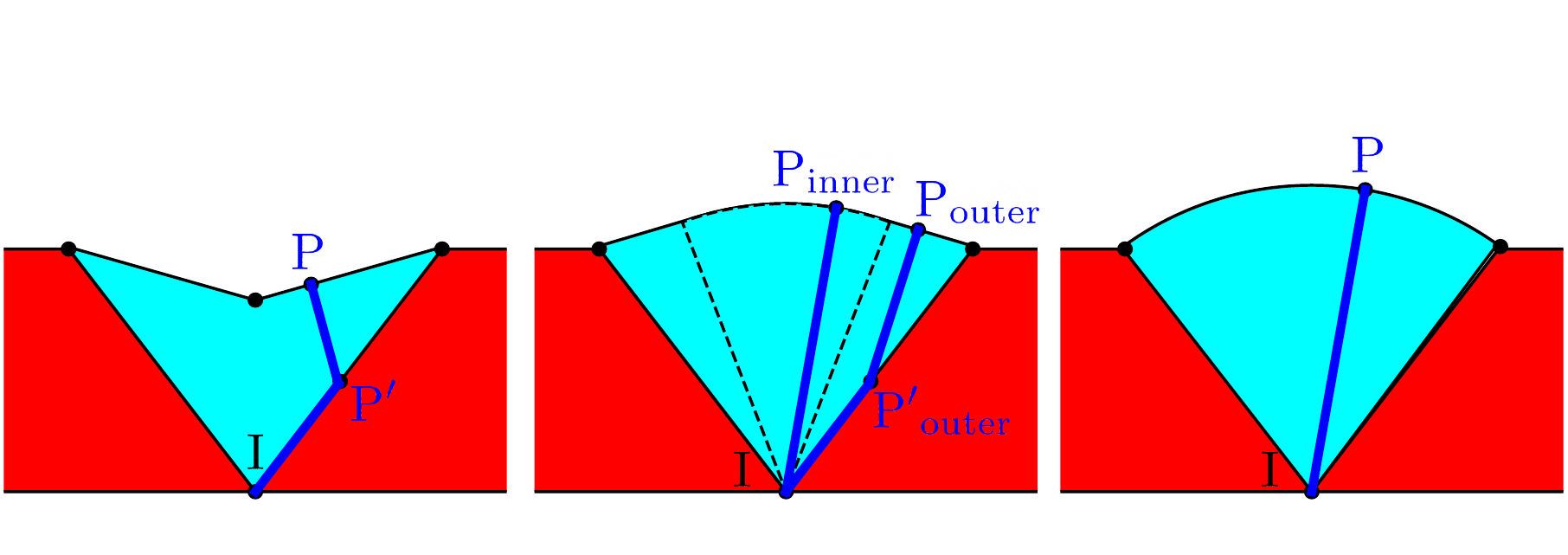}
	\centering
	\caption{Possible sector shapes for linear inoculation. Left:~$v_{\mathrm{\textsc{m}}} < v_{\mathrm{\textsc{w}}}$. The mutant sector emerging from point~$\mathrm{I}$ has a dented front. The front consists of two straight lines. The shortest-time path follows the sector boundary and also enters the sector interior. Middle:~$v_{\mathrm{\textsc{w}}} < v_{\mathrm{\textsc{m}}} < \sqrt{v_{\mathrm{\textsc{w}}}^2 + u^2}$. The mutant sector is a composite bulge. The front consists of two straight lines and an arc. To reach a point~$P_{outer}$ on straight part of the expansion front, the shortest-time path first follows the sector boundary before entering the sector interior. To reach a point~$P_{inner}$ on the arc, the shortest-time path follows a straight line from~$\mathrm{I}$ to~$P_{inner}$. Right: $v_{\mathrm{\textsc{m}}} > \sqrt{v_{\mathrm{\textsc{w}}}^2 + u^2}$. The front is an arc. To reach a point~$P$ on the front, the shortest-time path follows a straight line from~$\mathrm{I}$ to~$P$. }\label{fig:g3}
\end{figure}

The remaining possibility is the hybrid of the two cases considered so far. Depending on how far~$P$ is from the sector boundary, the quickest path from~$P$ to~$I$ may be either the direct or the indirect connection. We find that the front around~$x=0$ is a semicircle of radius~$v_{\mathrm{\textsc{m}}}t$, but it is a straight line near the sector boundaries. The two segments joint smoothly. The angular half-width of the central arc,~$\phi_{\mathrm{transition}}$, and the slope of the linear segment~(see Fig.~\ref{fig:g3}) are given by

\begin{align}
&\phi_{\mathrm{transition}} = \arctan\left( \frac {u v_{\mathrm{\textsc{m}}} - v_{\mathrm{\textsc{w}}} \sqrt{v_{\mathrm{\textsc{w}}}^2 + u^2 - v_{\mathrm{\textsc{m}}}^2}}{ v_{\mathrm{\textsc{m}}} v_{\mathrm{\textsc{w}}} + u\sqrt{v_{\mathrm{\textsc{w}}}^2 + u^2 - v_{\mathrm{\textsc{m}}}^2}} \right),\\
&{\rm slope} = -  \frac {u v_{\mathrm{\textsc{m}}} - v_{\mathrm{\textsc{w}}} \sqrt{v_{\mathrm{\textsc{w}}}^2 + u^2 - v_{\mathrm{\textsc{m}}}^2}}{ v_{\mathrm{\textsc{m}}} v_{\mathrm{\textsc{w}}} + u\sqrt{v_{\mathrm{\textsc{w}}}^2 + u^2 - v_{\mathrm{\textsc{m}}}^2}}. 
\end{align}

\subsection{Circular inoculation}

We assume that the expansion starts at $t=0$ when wildtype colony fills the circle with radius $r \le r_0$, and the mutant is present only at $I = (r_0,0)$ in polar coordinates.\\

\subsubsection{Sector boundary}
The boundary between the mutant and the wild type moves with linear velocity $u$ along the front.In polar coordinates, the position of the sector boundary~$(r_{\mathrm{\textsc{b}}}, \phi_{\mathrm{\textsc{b}}})$ then obeys the following equation
\begin{equation}
    \frac{d\phi_{\mathrm{\textsc{b}}}}{dt} = \frac{u}{r_{\mathrm{\textsc{b}}}}.
\end{equation}
We can eliminate time by using $dr_{\mathrm{\textsc{b}}}/dt = v_{\mathrm{\textsc{w}}}$ to obtain
\begin{equation}
    \phi_{\mathrm{\textsc{b}}}(r_{\mathrm{\textsc{b}}}) = \frac{u}{v_{\mathrm{\textsc{w}}}}\ln(\frac{r_{\mathrm{\textsc{b}}}}{r_0}).
\end{equation}

We also find that the length of boundary at time $t$ is $\sqrt{v_{\mathrm{\textsc{w}}}^2 + u^2}t$, and thus 
\begin{equation}
    v_{\mathrm{\textsc{b}}} = \sqrt{v_{\mathrm{\textsc{w}}}^2 + u^2}
\end{equation}
just as in the linear case.

\subsubsection{$v_{\mathrm{\textsc{m}}} < v_{\mathrm{\textsc{w}}}$}

Let us consider a point~$P = (r_p,\phi_p)$ on a mutant patch with $\phi_p>0$ for simplicity. \\
As described before, we first find~$T_{PP'I}$ by minimizing $\frac{|PP'|}{v_{\mathrm{\textsc{m}}} } + \frac{|\overset{\Huge\frown}{P'I}|}{v_{\mathrm{\textsc{w}}} }$ over points $P'$ on the sector boundary. The point $P' = (r_{P'}, \phi_{P'})$ should satisfy two equations: \\
\begin{equation}
   \phi_{P'}(r_{P'}) = \frac{u}{v_{\mathrm{\textsc{w}}}}\ln(\frac{r_{P'}}{r_0}),
\end{equation}
\begin{equation}
    \frac{d}{ dr_{P'}} \left( \frac{r_{P'} - r_0}{v_{\mathrm{\textsc{w}}} } + \frac{ \sqrt{(r_p \cos \phi_p - r_{P'} \cos \phi_{P'})^2 + (r_p \sin \phi_{P'} - r_{P'} \sin \phi_{P'})^2} }{ v_{\mathrm{\textsc{m}}} } \right) = 0.
\end{equation}

Here, the first equation constrains $P'$ to be on the sector boundary, and the second equation minimizes~$T_{PP'I}$ over $P'$. Since there are two unknowns and two equations, we can solve for $(r_{P'}, \phi_{P'})$. The solution is conveniently written in an implicit form:
\begin{equation}
\begin{aligned}
    &\frac{ r_{P'} \sin \phi_{P'} - r_p \sin \phi_p }{ r_{P'} \cos \phi_{P'} - r_p \cos \phi_p } = - \tan(\theta-\phi_{P'}), \\
    &\theta = \arctan\left( \frac {u v_{\mathrm{\textsc{m}}} - v_{\mathrm{\textsc{w}}} \sqrt{v_{\mathrm{\textsc{w}}}^2 + u^2 - v_{\mathrm{\textsc{m}}}^2}}{ v_{\mathrm{\textsc{m}}} v_{\mathrm{\textsc{w}}} + u\sqrt{v_{\mathrm{\textsc{w}}}^2 + u^2 - v_{\mathrm{\textsc{m}}}^2}} \right).
\end{aligned}
\end{equation}
This tells that $\overline{PP'}$ is parallel to $(1, \theta - \phi_{P'}) $; the angle between $\overline{PP'}$ and $\overline{P'O}$ is a constant $\theta$ independent of $r_p, \phi_p$. Note that $\theta > 0$ for $v_{\mathrm{\textsc{m}}} < v_{\mathrm{\textsc{w}}}$, and thereby every point $P$ on mutant front with $\phi_p$ has its corresponding $P'$ on sector boundary $\overset{\Huge\frown}{IB}$.\\

The next step toward identifying the front position at time~$T$ is to find all points~$P$ such that $T_{PP'I} = T$. Using the mapping between $P$ and $P'$ described above, we find $P$ by first moving along sector boundary and then moving in a straight line parallel to $(1, \theta - \phi_{P'})$. By varying the time $t'$ spent along the sector boundary while keeping the total time $T$ fixed, we obtain a parametric expression for $P(T) = (x_p(T), y_p(T) )$ in Cartesian coordinates:

\begin{equation}
\begin{aligned}
     &x_p(T;t') = (v_{\mathrm{\textsc{w}}} t'+r_0) \sin(\frac{u}{v_{\mathrm{\textsc{w}}}} \ln(\frac{r_0 + v_{\mathrm{\textsc{w}}} t'}{r_0})  )+ v_{\mathrm{\textsc{m}}} (T-t') \sin(\frac{u}{ v_{\mathrm{\textsc{w}}}} \ln(\frac{r_0 + v_{\mathrm{\textsc{w}}} t'}{r_0}) - \theta ),\\
     &y_p(T;t') = (v_{\mathrm{\textsc{w}}} t'+r_0) \cos(\frac{u}{v_{\mathrm{\textsc{w}}}} \ln(\frac{r_0 + v_{\mathrm{\textsc{w}}} t'}{r_0})  )+ v_{\mathrm{\textsc{m}}} (T-t') \cos(\frac{u}{ v_{\mathrm{\textsc{w}}}} \ln(\frac{r_0 + v_{\mathrm{\textsc{w}}} t'}{r_0}) - \theta ).
     \label{implicit}
\end{aligned}
\end{equation}

It is also possible to get a non-parametric, explicit expression by solving an equivalent partial differential equation using the method of characteristics:
\begin{equation}
\begin{aligned}
    & \phi_p(t,r) = \frac{u}{v\w}\ln\left(1 + \frac{v\w t}{r_0}\right) + F\left( \frac{r}{r_0 + v\w t} \right) - F(1),\quad \mathrm{where} \\
    & F(\rho) = \frac{u}{2v\w}\ln \left( (\rho^2 v\w^2 -v\m^2) \frac{ \sqrt{\rho^2 - \frac{v\m^2}{v\w^2 + u^2} } - \frac{u v\m}{v\w \sqrt{v\w^2 + u^2} } }{\sqrt{\rho^2 - \frac{v\m^2}{v\w^2 + u^2} } + \frac{u v\m}{v\w \sqrt{v\w^2 + u^2} } } \right) \\&+ \arctan \left( \frac{\sqrt{v\w^2 + u^2}}{v\m}\sqrt{\rho^2 - \frac{v\m^2}{v\w^2 + u^2}} \right).
    \label{explicit}
\end{aligned}
\end{equation}

\subsubsection{$v_{\mathrm{\textsc{m}}} > v_{\mathrm{\textsc{w}}}$}
In this regime, $\theta < 0$ and thereby some points $P$ on the mutant front do not have a corresponding $P'$ on the sector boundary. In other words, the straight path $\overline{IP}$ takes the shortest time. We find that, when~$P$ is near the top of the bulge, the minimal path is a straight line $\overline{IP}$ while, When~$P$ is further from the top, the minimal path is a straight line $\overline{P'P}$ followed by a curved path $\overset{\Huge\frown}{IP'}$ along the sector boundary.

Note that the straight path is tilted by a fixed angle $\theta$ from $\overline{OP'}$, pointing inwards to the center of the sector compared to the tangent line except when $v_{\mathrm{\textsc{m}}} = \sqrt{v_{\mathrm{\textsc{w}}}^2 + u^2}$. In the latter case, $\theta = -\arctan\left(\sqrt{ \frac{v_{\mathrm{\textsc{m}}}^2}{v_{\mathrm{\textsc{w}}}^2}-1} \right)$, and the straight path is tangent to the sector boundary, as described in~\cite{korolev:sectors}.  \\

The boundary between the two regions of the front lies angle~$\phi_{\mathrm{transition}}$ way from the center. This angle is given by
\begin{equation}
    \phi_{\mathrm{transition}} = \arctan\left( \frac {u v_{\mathrm{\textsc{m}}} - v_{\mathrm{\textsc{w}}} \sqrt{v_{\mathrm{\textsc{w}}}^2 + u^2 - v_{\mathrm{\textsc{m}}}^2}}{ v_{\mathrm{\textsc{m}}} v_{\mathrm{\textsc{w}}} + u\sqrt{v_{\mathrm{\textsc{w}}}^2 + u^2 - v_{\mathrm{\textsc{m}}}^2}} \right).
\end{equation}

Thus, the bulge is an arc of a circle near the center and is described by Eq.~(\ref{implicit}) near the sector boundary.

\clearpage
\section{Dispersal without carrying capacity}
\label{sec:const}
\begin{figure}[ht!]
\centering
\includegraphics[width=\textwidth]{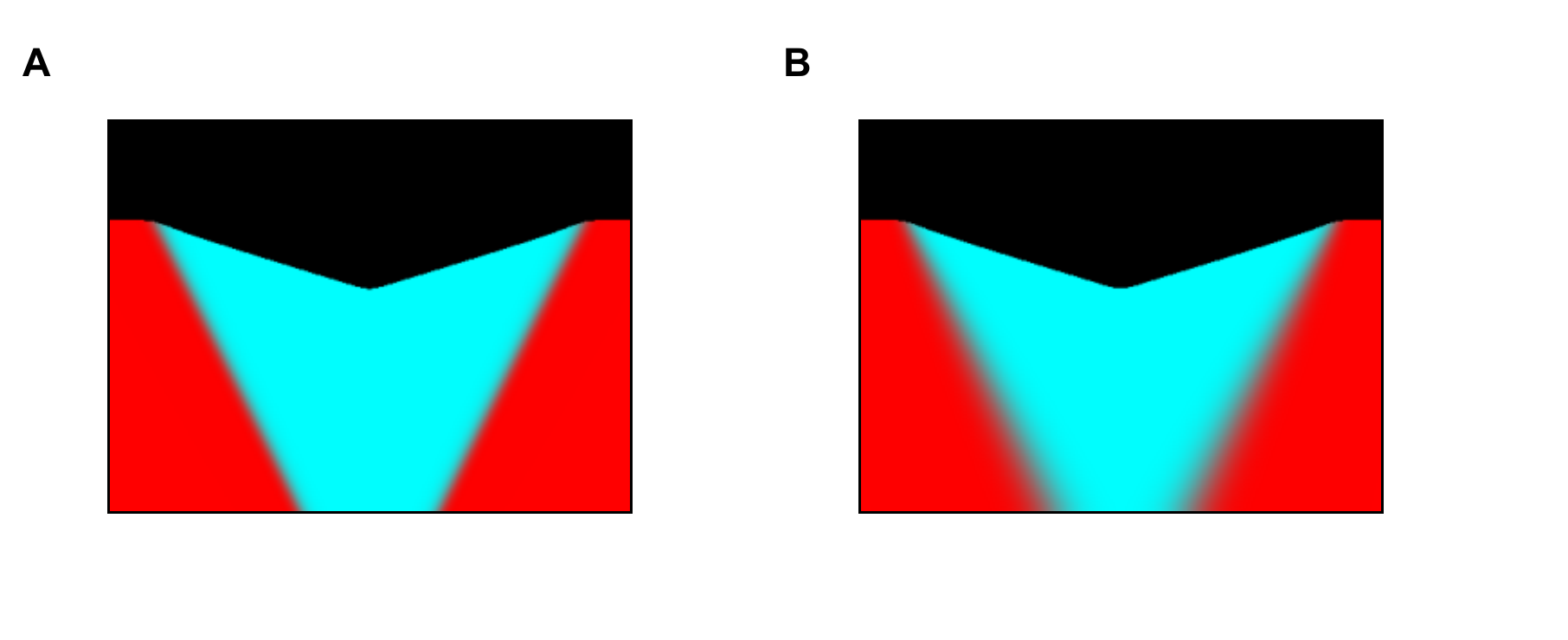}
\caption{Dented front in a model with density-independent dispersal. Formation of dented sectors in simulations with different models of dispersal. (A)~The model from the main text Eq.~\ref{rd}. (B) A model with density-independent dispersal Eq.~\ref{nocc}. Note the blurry sector boundaries due to continued intermixing after growth ceases behind the front.}\label{fig:si9}
\end{figure}

In the main text, we considered two mechanistic models that produce all possible sector shapes. For both models, we assumed that the dispersal term has a factor of $(1-n\w-n\m)$ so that the dispersal ceases when population reaches the carrying capacity. Without the carrying capacity factor, any spatial patterns should eventually vanish because the populations continue to intermix behind the expanding front. Accordingly, sectors exist only in the transient timescale between expansion and diffusion. Nevertheless, the $(1-n\w-n\m)$ factor does not affect the ratios between three velocities $v\w$, $v\m$ and $u$, and since these ratios determine the sector shape in geometric theory, we expect that the absence of the $(1-n\w-n\m)$ factor does not affect the sector shape observed in transient timescales. To verify this idea, we simulated a microscopic model without carrying capacity on diffusion:
\begin{equation}
\begin{aligned}
& \partial_t n\w =  D\nabla^2  n\w  + r\left(1 - \alpha \frac{n\m}{n\w + n\m} \right) n\w (1-n\w-n\m),\\
& \partial_t n\m = D\nabla^2  n\m  + r\left(1 - s + \alpha \frac{n\w}{n\w + n\m}\right) n\m (1-n\w-n\m).
\end{aligned}
\label{nocc}
\end{equation}
The simulation demonstrated that the sector shape was not affected by carrying capacity factor from dispersal (Fig.~\ref{fig:si9}). The sector boundaries were blurred by the nonzero dispersal behind the front, but the overall shape of the sector remained the same.
\clearpage
\section{Nonspatial limit for mechanistic models} \label{sec:ode}
In the main text, we considered two mechanistic models that produce all possible sector shapes. Here, we analyze these models in the nonspatial, i.e. well-mixed, limit, which describes local competition.

\subsection{Cheater-cooperator model}

The model reads

\begin{equation}
\begin{aligned}
& \partial_t n\w =  \left(D\nabla^2  n\w  + r\left(1 - \alpha \frac{n\m}{n\w + n\m} \right) n\w\right) (1-n\w-n\m),\\
& \partial_t n\m = \left(D\nabla^2  n\m  + r\left(1 - s + \alpha \frac{n\w}{n\w + n\m} \right) n\m\right) (1-n\w-n\m).
\end{aligned}
\label{ode1}
\end{equation}

In the well-mixed limit, the partial differential equations above reduce to a set of ordinary differential equations:

\begin{equation}
\begin{aligned}
& \frac{d n\w }{ dt} =   r\left(1 - \alpha \frac{n\m}{n\w + n\m} \right) n\w (1-n\w-n\m),\\
& \frac{d n\m }{ dt} =  r\left(1 - s + \alpha \frac{n\w}{n\w + n\m} \right) n\m (1-n\w-n\m).
\end{aligned}
\label{ode2}
\end{equation}

We only consider $s<1$ and $-1 < \alpha < 1$ and assume that initial populations densities are positive and their sum is below the carrying capacity. With these assumptions, it is clear that the population densities remain positive for any~$t\ge0$ since~$\frac{dn\w}{dt}$ and~$\frac{dn\m}{dt}$ are positive. The monotonic increase of the population densities also ensures that~$\lim_{t\to \infty} n\w+n\m = 1$ because both time derivatives switch sign when $n\w + n\m $ exceeds unity. In fact, it follows directly from~Eqs.~(\ref{ode2}) that any pair of positive~$n\w$ and~$n\m$ that sum up to one is a fixed point.

This line of fixed points is a direct consequence of our assumption that population dynamics are frozen behind the front. In a generic Lotka-Volterra system, the differences in the competitive abilities at high population densities would break this degeneracy and lead to the takeover by one of the types~(stable coexistence is also possible)~\cite{murray2002mathmatical, korolev:sectors}. Microbial populations however grow only until the nutrients are exhausted, and the two types could, therefore, remain at an arbitrary ratio once the growth ceases.

Further insights into the behavior of Eq.~(\ref{ode2}) can be derived from its first integral~(a conserved quantity), which we obtain by dividing the two equations:

\begin{equation}
    \frac{d n\w}{ d n\m} = \frac{\left(1 - \alpha \frac{n\m}{n\w + n\m} \right) n\w}{\left(1 - s + \alpha \frac{n\w}{n\w + n\m} \right) n\m}.
\label{ode3}
\end{equation}

The equation above can be integrated after both sides are multiplied by~$dn\m(1-s+\alpha n\w/(n\w+n\m))/n\w$. This procedure yields the following conserved quantity:

\begin{equation}
    C=\frac{ (n\w+n\m)^\alpha n\w^{1-s} }{n\m},
\label{ode4}
\end{equation}

which we can use to understand the temporal dynamics of the two types. It is convenient to recast Eq.~(\ref{ode4}) in terms of total population density~$n=n\w + n\m$ and mutant frequency~$f=n\m/n$:

\begin{equation}
\frac{f}{(1-f)^{1-s}} = \frac{n^{\alpha-s}}{C}.
\end{equation}

The left-hand side is a monotonically increasing function of~$f$, and the right hand-side is a monotonic function of~$n$, which is increasing for~$\alpha>s$ and decreasing otherwise. Thus,~$f$ increases with~$n$ for~$\alpha>s$ and decreases for~$\alpha<s$. Since~$n$ is always increasing~(assuming it is less than one initially), we conclude that the relative abundance of the mutant increases when~$\alpha>s$ and decreases otherwise. Numerical simulations confirm this conclusion; see Fig.~\ref{fig:si8}A.

In the spatial model, $u = \sqrt{\alpha - s}$, so the mutant can invade only when $s < \alpha$, which is consistent with the local well-mixed competition that we just described.

\subsection{Growth-dispersal tradeoff model}

The well-mixed limit for the growth-dispersal tradeoff model reads

\begin{equation}
\begin{aligned}
& \frac{ n\w }{ dt} =   r n\w (1-n\w-n\m),\\
& \frac{ n\m }{ dt} =  r(1+s) n\m (1-n\w-n\m).
\end{aligned}
\label{ode5}
\end{equation}

The qualitative behavior of this system of equations is identical to that of the cheater-cooperator model. Any population below the carrying capacity with positive densities of the two types evolves to one of the neutral fixed points on~$n\m+n\w=1$. The change of the mutant fraction can be determined from the following first integral

\begin{equation}
    \frac{ n\w^{1+s} }{n\m} = n^s\frac{(1-f)^{1+s}}{f} = C.
\label{ode7}
\end{equation}

\begin{figure}[ht!]
\centering
\includegraphics[width=\textwidth]{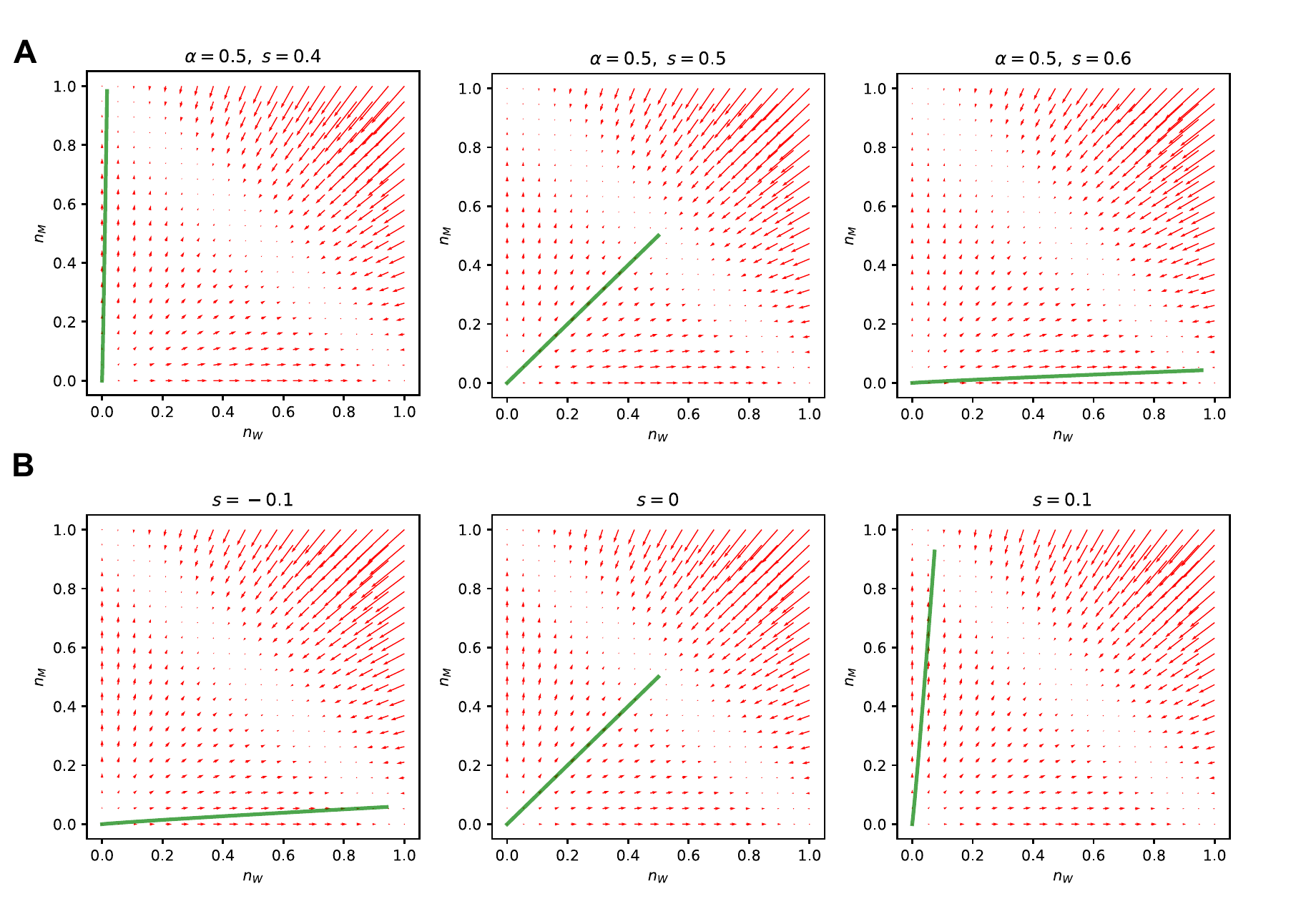}
\caption{Phase portraits of ODE dynamics. In each panel, red arrows represent $(dn\w/dt, dn\m/dt)$ and green curve shows the trajectory from small initial population $(n\w,n\m)=(10^{-12},10^{-12})$. (A)~Phase portraits for cheater-cooperator interaction model. (B)~Phase portraits for growth-dispersal tradeoff model. }\label{fig:si8}
\end{figure}

The analysis, identical to the one we just described, shows that the frequency of the mutant increases as long as~$s>0$. This is consistent both with the expansion velocity~$u=2\sqrt{Ds}$ and numerical simulations~(Fig.~\ref{fig:si8}).

\clearpage

\printbibliography
\end{document}